\documentclass[oribibl,orivec]{llncs}

\usepackage{alltt}
\usepackage{amsmath}
\usepackage{amssymb}
\usepackage{semantic} 
\usepackage{stmaryrd}
\usepackage{bm}
\usepackage{color}
\usepackage{xypic}
\usepackage{graphics}
\usepackage{graphicx}
\usepackage{listings}
\usepackage{fancybox}
\usepackage{flushend}
\usepackage{xcolor}
\usepackage{mathpartir}

\colorlet{lprolog}{blue!70!black}
\colorlet{abellatop}{blue!70!green}
\colorlet{abellatac}{orange!30!black}
\colorlet{abellabad}{red!80!yellow}

\lstdefinelanguage{lprolog}{%
  alsoletter={-},
  classoffset=0,%
  morekeywords={sig,module,type,kind,pi,sigma,end},%
  keywordstyle=\color{lprolog},%
  classoffset=0,%
  otherkeywords={:-,=>,<=,\&},%
  sensitive=true,%
  morestring=[bd]",%
  morecomment=[l]\%,%
  morecomment=[n]{/*}{*/},%
}

\lstdefinelanguage{abella}[]{lprolog}{%
  alsoletter={-},
  classoffset=1,%
  morekeywords={Close,CoDefine,Define,Kind,Query,Quit,Specification,
    Set,Split,Theorem,Type,Undo,by,as,prop,true,false,forall,exists,nabla},%
  keywordstyle=\color{abellatop},%
  classoffset=2,%
  morekeywords={abbrev,apply,backchain,case,coinduction,cut,
    induction,inst,intros,monotone,on,permute,rename,left,right,witness,
    search,split,to,unabbrev,unfold,assert,with},%
  keywordstyle=\color{abellatac},%
  classoffset=3,%
  morekeywords={undo,abort,skip,clear},%
  keywordstyle=\color{abellabad}\underbar,%
  classoffset=0,%
}

\usepackage{mathtools}
\usepackage{relsize}

\newcommand{\exh}{\stackrel{\texttt{exh}}{\sim}}

\newcommand{\caseAbbr}[3]{\key{case}\,(x)\,#1\,#2\,#3}

\newcommand{\lambdafull}{\texttt{Fpl}}

\newcommand{\lpGamma}{\mathbf{\Gamma}}

\newcommand{\lp}[1]{\blue{\textsf{\textbf{#1}}}}

\newcommand{\multistepTL}{\lp{$\stepShort^{*}$}}
\newcommand{\typeofTL}{\lp{$\vdash$}}
\newcommand{\stepTL}{\lp{$\stepShort$}}
\newcommand{\valueTL}{\blue{\textsf{\textbf{value}}}}
\newcommand{\errorTL}{\blue{\textsf{\textbf{error}}}}
\newcommand{\valuesTL}{\mathit{V}}
\newcommand{\contextsTL}{\mathit{ctx}}

\newcommand{\ninference}[3]{\inferrule[(#1)]{#2}{#3}}

\newcommand{\certifier}{TypeSoundnessCertifier}

\newcommand{\defSup}{^{\text{d}}}
\newcommand{\typSup}{^{\text{t}}}
\newcommand{\redSup}{^{\text{r}}}
\newcommand{\valueSup}{^{\textrm{v}}}

\newcommand{\tsSub}{_{\textrm{ts}}}
\newcommand{\prgSub}{_{\textrm{prg}}}
\newcommand{\preSub}{_{\textrm{pre}}}
\newcommand{\chrSub}{_{\textrm{symb}}}
\newcommand{\defSub}{_{\textrm{def}}}
\newcommand{\typSub}{_{\textrm{typ}}}

\newcommand{\entSub}{_{\textrm{ent}}}

\newcommand{\ba}{\begin{array}}
\newcommand{\ea}{\end{array}}

\newenvironment{syntax}{\[\ba{l@{\;\;}lcl}}{\ea\]}

\newcommand{\dotspace}{.\,}

\definecolor{ShadowColor}{RGB}{30,150,190}

\makeatletter
\newcommand\Cshadowbox{\VerbBox\@Cshadowbox}
\def\@Cshadowbox#1{%
  \setbox\@fancybox\hbox{\fbox{#1}}%
  \leavevmode\vbox{%
    \offinterlineskip
    \dimen@=\shadowsize
    \advance\dimen@ .5\fboxrule
    \hbox{\copy\@fancybox\kern.5\fboxrule\lower\shadowsize\hbox{%
      \color{ShadowColor}\vrule \@height\ht\@fancybox \@depth\dp\@fancybox \@width\dimen@}}%
    \vskip\dimexpr-\dimen@+0.5\fboxrule\relax
    \moveright\shadowsize\vbox{%
      \color{ShadowColor}\hrule \@width\wd\@fancybox \@height\dimen@}}}
\makeatother

\newcommand{\typeOf}{\vdash}
\newcommand{\step}{\longrightarrow}
\newcommand{\stepShort}{\rightarrow}

\newcommand{\ifExp}[3]{\key{if}\app #1\app \key{then}\app #2\app \key{else}\app #3}

\newcommand{\caseE}[3]{\key{case}\,#1\,\key{of}\,\key{inl}\,x\Rightarrow\,#2\,| \,\key{inr}\,y\Rightarrow \,#3}

\newcommand{\proptype}{\lp{o}}

\newcommand{\type}{\vdash\,}

\newcommand{\equalUnique}{\stackrel{\texttt{unq}}{=}}

\definecolor{lightblue}{rgb}{0.25,0.25,1}

\newcommand{\blue}[1]{\textcolor{lightblue}{#1}}
\newcommand{\redd}[1]{\textcolor{red}{#1}}

\definecolor{lightgray}{gray}{0.9}
\newcommand{\HI}[1]{\colorbox{lightgray}{#1}}

\newcommand{\lpsyntax}[1]{\ensuremath{\mathit{#1}}}
\newcommand{\key}[1]{\ensuremath{\mathtt{#1}}}

\newcommand{\Int}{\key{Int}}
\newcommand{\Bool}{\key{Bool}}

\newcommand{\lam}[1]{\lambda #1 \dotspace}
\newcommand{\app}{\;}
\newcommand{\of}{{:}}

\newcommand{\op}{\mathit{op}}

\newcommand{\tagsc}[1]{\tag{\textsc{#1}}}

\definecolor{lightgray}{gray}{0.9}

\newcommand{\zero}{{\key{z}}}
\newcommand{\s}{{\key{succ}}}
\newcommand{\pred}{{\key{pred}}}

\newcommand{\true}{{\key{true}}}
\newcommand{\false}{{\key{false}}}
\newcommand{\List}{{\key{List}}}

\newcommand{\try}[2]{\key{try} \app #1 \app \key{with} \app #2}
\newcommand{\tryExp}{\key{try}}
\newcommand{\letrec}[3]{\key{letrec} \app #1=#2 \app \key{in} \app #3}
\newcommand{\letrecExp}{\key{letrec}}

\newcommand{\nil}{{\key{nil}}}
\newcommand{\cons}{{\key{cons}}}
\newcommand{\head}{{\key{head}}}
\newcommand{\tail}{{\key{tail}}}
\newcommand{\fold}{{\key{fold}}}
\newcommand{\unfold}{{\key{unfold}}}
\newcommand{\myRaise}{{\key{raise}}}
\newcommand{\fix}{{\key{fix}}}

\newcommand{\typedLanguage}{\mathbb{T}}
\newcommand{\Ldl}{\mathcal{L}}

\newcommand{\field}[1]{\text{\scriptsize{#1}}}

\newcommand{\redSub}{_{\text{red}} }

\newcommand{\emptyLdl}[1]{\Ldl^{\epsilon}}

\newcommand{\V}{\blue{\key{(V)}}}
\newcommand{\E}{\blue{\key{(E)}}}
\newcommand{\Ss}{\blue{\key{(S)}}}
\newcommand{\Pp}{\blue{\key{(P)}}}

\begin{document}

%


\title{Well-Typed Languages are Sound}

\author{Matteo Cimini\inst{1} \and Dale Miller\inst{2} \and Jeremy G. Siek\inst{1} }
\institute{Indiana University Bloomington\\ \and
  INRIA and LIX/\'{E}cole Polytechnique}
\maketitle

  \renewcommand{\baselinestretch}{0.91}%

\begin{abstract}
Type soundness is an important property of modern programming 
languages. In this paper we explore the idea that 
\emph{well-typed languages are sound}: the idea 
that the appropriate typing discipline 
over language specifications guarantees that the language is type sound. 
We instantiate this idea for a certain class of languages defined using small step
operational semantics by ensuring the progress 
and preservation theorems. 

Our first contribution is a syntactic discipline for organizing and 
restricting language specifications so that they automatically satisfy 
the progress theorem. 
This discipline is not novel but makes explicit 
the way expert language designers have been organizing a certain class 
of languages for long time. 
We give a formal account of this discipline by representing 
language specifications as (higher-order) logic
programs and by giving a meta type system over that collection of
formulas.
Our second contribution is an analogous methodology and meta type system for 
guaranteeing that languages satisfy the preservation theorem. 
Ultimately, we have proved that language specifications that conform to our meta type systems 
are guaranteed to be type sound. 
%

We have implemented these ideas in the \emph{TypeSoundnessCertifier}, 
a tool that takes language specifications in the form of logic programs and type checks them 
according to our meta type systems. 
For those languages that pass our type checker, 
our tool automatically produces a proof of type soundness that can be 
independently machine-checked by the Abella proof assistant. 
%
For those languages that fail our type checker, the tool pinpoints the design mistakes that hinder type soundness.
We have applied the \emph{TypeSoundnessCertifier} tool to a large 
number of programming languages, including those with recursive types, 
polymorphism, letrec, exceptions, lists and other common types and 
operators. 
\end{abstract}

\section{Introduction}
\label{sec:introduction}

Types and type systems play a fundamental role in programming
languages. They provide programmers with abstractions, documentation, and 
useful invariants. The run-time behavior of programs is
oftentimes a delicate and unpredictable matter.
However, through the use of types and good design choices, 
programming languages can often ensure that, during run-time, desirable
properties are maintained and unpleasant behaviors are eliminated.
Of all the
properties that we wish to establish for typed languages, type soundness is one of the
most important. Type soundness can be summarized with Robin Milner's
 slogan that says that \emph{well typed programs
  cannot go wrong}: that is, they cannot get stuck at run-time. 
  
In this paper we explore the idea that 
\emph{well-typed languages are sound}: the idea that the appropriate typing discipline 
over language specifications guarantees that the language is type sound. 

We instantiate this idea to a certain class of programming languages 
defined in small step operational semantics and we follow the approach of 
Wright and Felleisen. 
In their paper \emph{A Syntactic Approach to Type Soundness} \cite{wright94ic},
Wright and Felleisen offered an approach to proving type soundness
that has become a de facto standard and that relies on two
 key properties: the progress and type preservation theorems. 
%
Progress states that if a program is well-typed then it is either a
value, an error, or it performs a reduction. Type preservation states
that if a program has some type, a reduction step takes it to a
program that has the same type.

Our first contribution is a methodology for organizing and 
restricting language definitions so that they automatically satisfy 
the progress theorem. 
An important aspect of the methodology is the classification of the operators of the language at hand. 
For example, some operators are
\emph{constructors} that build \emph{values}, such as the
functional space constructor $\lambda x. e$
in the simply typed $\lambda$-calculus (STLC). 
Some other operators are 
\emph{eliminators}: e.g., application. 
Other kinds of operators are \emph{derived operators} (such as \key{letrec}), \emph{errors} and \emph{error handlers}. 
The overall discipline is descriptive and simply
resembles the way programming languages have been defined for a long time. 
For example, among other restrictions, the discipline imposes that
eliminators must have reduction rules for every value allowed by the
type of their argument and that those arguments that need be evaluated
to a value must be set as evaluation contexts.
%
%
%
%

In our formalization of this descriptive methodology, 
we represent language specifications using logic programs. 
This is a convenient choice since, as has been argued long ago by
Sch\"urmann and Pfenning \cite{schurmann98cade}, such specifics are
executable, correspond closely with pen \& paper specifications,
and have a formal semantics that can be the subject of proofs.
We give a meta type system over language specifications that directly 
imposes the mentioned discipline.
To make an example, the $\beta$ rule $(\lambda x. e)\app v  \step
e[v/x]$ can be type checked in the following way. (The application
operator is named here as \textit{app}.)
\[
\inference
	{ 	  
	\Gamma(\blue{app}) = \text{elim}\app \blue{\rightarrow} & \Gamma(\blue{\lambda}) = \text{value}\app \blue{\rightarrow} \app \emptyset\\
	 \{1, 2\} \subseteq \contextsTL(\blue{app}) 
           	} 
	{ \contextsTL \mid \Gamma \typeOf \blue{(app \app
           \HI{$(\lambda x.\app e)$} \app v) \app\stepTL\app e[v/x]}: \blue{app} : \text{eliminates}\app \blue{\lambda}
           }
\]
That is, the rule is well-typed because the application is an eliminator of the function type and its \emph{eliminated argument}, high-lighted, is a value of the function type. Moreover, the arguments at positions $1$ and $2$ are set as evaluation contexts for the application. 
The meta typing rule assigns the type ``$app : \text{eliminates}\app \lambda$'' so that the type system has a means to check later whether $app$ eliminates all the values of $\rightarrow$, which, in this case, is just the function. 


\begin{figure*}[tbp]
{\scriptsize \textrm{stlc\_cbv.mod:}}
\begin{lstlisting}[numbers=left,language=lprolog]
module stlc_cbv.

typeOf (abs T1 E)  (arrow T1 T2) :- (pi x\ typeOf x T1 => typeOf (E x) T2).
typeOf (app E1 E2) T2            :- typeOf E1 (arrow T1 T2), typeOf E2 T1.
typeOf tt bool.
typeOf ff bool.
typeOf (if E1 E2 E3) T :- typeOf E1 bool, typeOf E2 T, typeOf E3 T.
step   (app (abs T E) V) (E V)    :- value V.
step (if tt E1 E2) E1.
step (if ff E1 E2) E2.
value  (abs T1 R2).
value tt.
value ff.

% context app E e 
% context app v E.
% context if E e e.
\end{lstlisting}
\caption{Example input of the \emph{\certifier}: file \textrm{stlc\_cbv.mod}. This is the formulation of STLC with the \key{if}  operator. }
\label{fig:output}
\end{figure*}

The type preservation theorem is not, generally speaking, ensured by a
discipline. However, typing is markedly happening. 
For the $\beta$ rule we have to ensure that the type of $(\lambda x. e)\app v$ is
the same as the type of $e[v/x]$. However, these are expressions with
variables and their types depend on the type assumptions on 
their variables: that is, they depend on $\Gamma$ of typing judgments. 
Ideally, we need to check that for all $\Gamma$, if $\Gamma \vdash
(\lambda x. e)\app v : T$ then $\Gamma \vdash e[v/x] : T$.
Such a
statement is prohibitive to check due to the quantification over all
$\Gamma$s. 
Nonetheless we are able to offer a methodology for type preservation.
The methodology fixes a \emph{symbolic
  type environment $\Gamma^{s}$} based on the information
extracted from the typing rules on which the expressions of $\beta$ rely
on. We take a practical approach by representing $\Gamma^{s}$ as a
conjunction of typing formulae and use entailment for checking that
the types of $(\lambda x. e)\app v$ and
$e[v/x]$ agree. For example, by inspecting the typing rules for
application and abstraction we build and check the formula
\[
  \typeOf \app v : T_1 \; \land \;
  (x : T_1 \typeOf \app e : T_2)~ \Rightarrow~
  \typeOf \app e[v/x] : T_2. 
\]
This approach fits naturally a type system formulation. 
Analogously to the case of progress, we devise a meta type system for 
languages that automatically satisfy the type preservation theorem. 

Ultimately, we have proved that languages that conform to our meta type systems satisfy both 
progress and type preservation. This validates the methodologies in this paper and, possibly, proves that
the invariants that language designers have been using for a long time are correct.
As a consequence of our results, language specifications that type
check successfully are guaranteed to be type sound: hence the slogan
``\emph{well-typed languages are sound}''.

Based on our results, we have implemented the \emph{\certifier} tool. 
%
The tool works with language specifications such as that in \key{stlc\_cbv.mod} of 
Figure~\ref{fig:output}. 
This file contains the formulation of the STLC with the \key{if} operator. 
The specification language is that of $\lambda$Prolog \cite{Miller:2012aa} augmented with
convenient context tags for declaratively specifying evaluation
contexts.  
The \emph{\certifier} tool can input this file and type check the language specification according to 
the meta type systems devised in this paper. 
If type checking succeeds, the tool automatically generates the
theorems and proofs for the progress, type preservation, and ultimately
type soundness theorems. 
These proofs are then machine-checked against an external proof assistant. 
In particular, we use the Abella \cite{baelde14jfr} proof
assistant  (which can load and reason with $\lambda$Prolog specifications) 
as a proof-checker for the proofs produced by \emph{\certifier}.
If type checking fails, the tool reports a meaningful error to the user. Were we to forget the tag at line 15 (\texttt{\% context app E e.}) of Figure~\ref{fig:output}, the \emph{\certifier} would reject the specification and 
tell the user that the first argument of the application must be an evaluation context. 
Were we to forget one of the reduction rules for \key{if}, say line 10, the type checking would fail reporting that this eliminator for \key{bool} does not eliminate \emph{all} the values of type \key{bool}. 

In summary, this paper makes the following contributions.

\smallskip$(1)$ We offer a complete methodology for ensuring the type soundness of languages (Sections \ref{sec:classification}, \ref{sec:progress:methodology} and \ref{sec:typePreservation}). 
The target of our methodology is a class of languages that is based on constructors/eliminators and errors/error handlers, that is common in programming languages design. This class of languages is fairly expressive and accommodates modern features such as recursive types, polymorphism, and exceptions. 

\smallskip $(2)$ We formulate the methodology as a meta type system over language specifications (Section \ref{sec:typeLanguagesAsLogicPrograms} and \ref{sec:typeSystemForTypedLanguages}). 
We have proved that our meta type system guarantees the type soundness 
of languages (Section \ref{sec:not-unsound}). This validates the common practice that language designers have been used
for long and demonstrates the idea that \emph{well-typed
  languages are sound}.

\smallskip $(3)$ We implemented the \emph{\certifier} tool that can
certify a language as being type sound or it can pinpoint design mistakes (Section \ref{sec:implemenation}). 
We have applied
our tool to the type checking of several languages, including variants of STLC, 
as well as its implicitly typed version,
with the following features: pairs,
\key{if\textendash then\textendash else}, lists, sums, unit, tuples,
\key{fix}, \key{let}, \key{letrec}, universal types, recursive types
and exceptions.  We have also considered different evaluation strategies among call-by-value, call-by-name
and a parallel reduction strategy, as well as lazy pairs, lazy lists
and lazy tuples.
In total, we have type checked $103$ type
sound languages. 
\emph{\certifier} has automatically generated proof of progress,
preservation and type soundness for each of the type checked languages and these proofs
have been independently checked by an external proof checker.
This gives us high confidence in our type systems. 

The \emph{\certifier} tool can be found at the following repository:

\centerline{{\texttt{https://github.com/mcimini/TypeSoundnessCertifier}}}

In the next section, we briefly review some terminology in the context of typed languages
that are defined in small step operational semantics.

\section{Typed Languages}
\label{sec:typedLanguagesInformally}


 Let us consider the 
language $\lambdafull$ defined in Figure
\ref{fig:LambdaFull}. This language is a fairly involved
programming language with integers, booleans, 
\key{if\textendash then\textendash else}, sums, lists, 
universal types, recursive types, \key{fix}, \key{letrec} and
exceptions.

Types and expressions are defined by a BNF grammar. Next, 
language designers decide which expressions constitute \emph{values}.
These are the possible results of successful computations. Similarly,
the language designer may define which expressions constitute
\emph{errors}, which are possible outcomes of computations when
they fail. 

The top part of Figure \ref{fig:LambdaFull:semantics} shows the type
system for $\lambdafull$. 
The type system is an inference rule system for judgements that, in this paper, have 
the form $\Gamma \typeOf e : T$. 
A term that is constructed with the application of a type constructor to distinct variables is called a
\emph{constructed type}. For example, $\List\app T$ and $T_1 \to T_2$ are constructed types. 
$\Int$ is a constructed type, as well, because it simply has arity $0$. 
Analogously, expressions like
$\key{fold}\app e$ and $\cons\app e_1\app e_2$ are \emph{constructed expressions}. 
Given a typing rule such as $\small{\inference
	{
	\Gamma \typeOf e_1 : T &
	\Gamma \typeOf e_2 : {\List\app T}
	 }
	{ \Gamma \typeOf \app {\cons \app e_1\app e_2}:  \HI{$\List\app T$}}}$ 
we say that the high-lighted ${\List\app T}$ is the \emph{assigned type}.

The bottom part of Figure \ref{fig:LambdaFull:semantics} defines the 
dynamic semantics of $\lambdafull$. It is defined by a series 
of \emph{reduction rules}. For a formula
$e\step e'$, $e$ is the \emph{source} and $e'$ is the \emph{target} of the reduction. 
In a reduction rule such as \textsc{(r-head-cons)}, i.e. $\small{{\head \app({{\cons \app v_1\app v_2}})}  \step {v_1}}$, we say that the first  argument of $\head$ is \emph{pattern-matched} against the constructed expression 
$(\cons \app v_1\app v_2)$.

The dynamic semantics of a language is also defined by its 
\emph{evaluation contexts}, which prescribe within which
context we allow reduction to take place. 
They are defined with the syntactic category 
$\text{Context}$ of Figure \ref{fig:LambdaFull}. 
For a context definition such as $\key{cons}\app E\app e$ we
say that the first argument of \key{cons} is \emph{contextual}.

Error contexts define which 
contexts are allowed to make the whole computation fail when we spot an
error.

\begin{figure}[tbp]
\scriptsize
\begin{syntax}
  \text{Types} & T & ::= & \Bool \mid \Int \mid \app T\to  T \mid 
  \List \app T \mid T + T\\
  &&& X \mid \forall X.T \mid \mu X.T\\
  \text{Expressions} & e & ::= & 
  \true \mid \false \mid \ifExp{e}{e}{e} \\
  &&& \mid \zero \mid \s\app e \mid \pred \app e \mid \key{isZero}\app
  e\\
  &&& \mid x \mid \lambda x.e\mid e\app e \\
  &&& \mid \nil \mid \cons\app e\app e\mid \\
&&& \mid \head \app e \mid  \tail \app e \mid \key{isNil}\app
  e\\
  &&& \mid \key{inl} \app e \mid \key{inr} \app e \mid \caseAbbr{e}{e}{e}\\
  &&& \mid \Lambda X. \app e \mid e\app [T]  \\
  &&& \mid \fold \app e \mid \unfold\app e  \\
  &&& \mid \fix\app e \mid \letrec{x}{e}{e}  \\
  &&& \mid \myRaise\app e \mid \try{e}{e}  \\
   \text{Values} & v & ::= & \true \mid \false  
\mid \zero \mid \s\app v   
\mid\lambda x.e 
 \mid\key{nil} \mid \key{cons}\app v\app v \\   
   &&& 
\mid \key{inr} \app v \mid \key{inl} \app v 
\mid \Lambda X. \app e 
\mid\key{fold}\app v \\
   \text{Errors} & \mathit{er} & ::= & \key{raise}\app v  \\
  \text{Contexts} & E & ::=  &  \ifExp{E}{e}{e}\\
   &&& \mid \s\app E\mid \pred \app E \mid \key{isZero}\app
  E\\
   &&& \mid E\app e\mid v\app E \\
   &&& \mid \key{cons}\app E\app e\mid\key{cons}\app v\app E 
  \mid
   \key{head}\app E \mid  \key{tail}\app E \mid \key{isNil}\app E\\
   &&& \mid \key{inl}\app E \mid \key{inr} \app E \mid \caseAbbr{E}{e}{e}\\
   &&& \mid E\app [T] \mid\key{fold}\app E\mid\key{unfold}\app E 
    \mid \fix\app E\mid \letrec{x}{E}{e}\\
   &&& \mid \myRaise\app E \mid\try{E}{e}\\ 
\end{syntax}

Error Contexts, $F$, are just Contexts but without the $(\try{E}{e})$ case.
\caption{The syntax of $\lambdafull$ contains a number of features
  that are all handled by our analysis.  This language is not
  \emph{minimal} since, for example, recursive types can define
  booleans and lists. $\caseAbbr{e}{e}{e}$ is short for $\caseE{e}{e}{e}$.}
\label{fig:LambdaFull}
\end{figure}

\begin{figure}[tbp]
\scriptsize
Type System  \hfill  \fbox{$\Gamma \vdash e : T$}
\begin{gather*}
{\Gamma,x:T\typeOf\kern -1pt x:T}\quad
{\Gamma \typeOf \app \true:\Bool}
\qquad 
{\Gamma \typeOf \app \false:\Bool}	
\\
\inference
	{\Gamma \typeOf \app e_1 : \Bool &
	\Gamma \typeOf \app e_2 : T &
	\Gamma \typeOf \app e_3 : T 
	} 
	{ \Gamma \typeOf \app\ifExp{e_1}{e_2}{e_3} : T}  \,\,\textsc{(t-if)}
\\
\ninference{t-z}{}{\Gamma \typeOf \zero:\Int}	
\quad 
\ninference{t-succ}
	{\Gamma \typeOf e:\Int 
	} 
	{ \Gamma \typeOf \s \app e: \Int}  
\quad 
\ninference{t-pred}
	{
	\Gamma \typeOf e : \Int 
	} 
	{ \Gamma \typeOf \pred \app e: \Int}  
\quad 
\ninference{t-iszero}
	{
	\Gamma \typeOf e : \Int 
	} 
	{ \Gamma \typeOf \key{isZero} \app e: \Bool}  
\\
\ninference{t-lambda}
	{\Gamma,x:T_1 \typeOf \app e : T_2} 
	{ \Gamma \typeOf \app \lambda x. e:  T_1\to T_2}
\qquad 
\ninference{t-app}
	{
	\Gamma \typeOf \app e_1 : T_1\to T_2 \\
	\Gamma \typeOf \app e_2 : T_1 	
	} 
	{ \Gamma \typeOf \app e_1\app e_2 : T_2}
\\
\ninference{t-nil}{}{
\Gamma\typeOf \app \nil: \List \app T}
\qquad
\ninference{t-cons}
	{
	\Gamma \typeOf e_1 : T \\ 
	\Gamma \typeOf e_2 : {\List\app T}
	 }
	{ \Gamma \typeOf \app {\cons \app e_1\app e_2}:  {\List\app T}}
\\
\ninference{t-head}
	{\Gamma \typeOf \app e : \List \app T} 
	{ \Gamma \typeOf \app \head \app e:  T}
\qquad 
\ninference{t-tail}
	{\Gamma \typeOf \app e : \List \app T} 
	{ \Gamma \typeOf \app \tail \app e:  \List\app T}
\quad 
\ninference{t-isnil}
	{
	\Gamma \typeOf e : {\List\app T} 
	} 
	{ \Gamma \typeOf \key{isNil} \app e: \Bool}  
\\
\inference
	{\Gamma \typeOf e:T_1 
	} 
	{ \Gamma \typeOf  \key{inl}\app e : T_1+ T_2}  \,\,{\small \textsc{(t-inl)}}
\quad 
\inference
	{\Gamma \typeOf e:T_2 
	} 
	{ \Gamma \typeOf  \key{inr} \app e : T_1+ T_2} \,\,{\small \textsc{(t-inr)}}
\\
\ninference{t-case}
{\Gamma \type e_1 : {T_1}+{T_2}\\
 \Gamma, x:{T_1}\type e_2: {T} \\
 \Gamma, x:{T_2}\type e_3: {T}}
{\Gamma \type (\caseE{e_1}{e_2}{e_3}) :
  T}
\\
\ninference{t-abst}
{\Gamma,X \typeOf \app e : T} 
          { \Gamma \typeOf \app \Lambda X.  e :  \forall X.T}
\quad\quad
\ninference{t-appt}
{ \Gamma\typeOf \app e: \forall X.T_2} 
          { \Gamma\typeOf \app ( e\app [T_1]) : T_2[T_1/X]}
\\
\inference
	{\Gamma\typeOf \app e: T[\mu X. T/X]} 
	{ \Gamma\typeOf \app \fold \app e: \mu X. T}\,\,{\small \textsc{(t-fold)}}
\quad\quad
\inference
	{\Gamma \typeOf \app e: \mu X. T} 
	{\Gamma \typeOf \app \unfold \app e : T[\mu X. T/X]}\,\,{\small \textsc{(t-unfold)}}
\\
\ninference{t-fix}
	{\Gamma \typeOf e : T \to T}
	{\Gamma \typeOf \app \fix \app e : T}
\qquad 
\ninference{t-letrec}
	{
	\Gamma, x:T_1 \typeOf e_1 : T_1 \\
	\Gamma, x:T_1 \typeOf e_2 : T_2 
	 }
	{\Gamma \type\app  \letrec{x}{e_1}{e_2} : T_2}
\\
\ninference{t-raise}
	{\Gamma \typeOf e : \Int }
	{\Gamma \typeOf \app \myRaise \app e : T}
\qquad 
\ninference{t-try}
	{
	\Gamma \typeOf e_1 : T \\
	\Gamma \typeOf e_2 : \Int \to T
	 }
	{\Gamma \type\app  \try{e_1}{e_2} : T}
\end{gather*}
Dynamic Semantics  \hfill \fbox{$e\step e$}
\begin{align*}
	\ifExp{\true}{e_1}{e_2} & \step  e_1     	\label{r-if-true}\tagsc{r-if-true}\\[-5pt]
	\ifExp{\false}{e_1}{e_2} &  \step  e_2    	\label{r-if-false}\tagsc{r-if-false}\\[-5pt]
	\pred\app \zero & \step  \myRaise \app \zero    \label{r-pred-zero}\tagsc{r-pred-zero}\\[-5pt]
	\pred\app (\s \app e) & \step  e     	        \label{r-pred-succ}\tagsc{r-pred-succ}\\[-5pt]
	\key{isZero}\app \zero & \step  \true     	\label{r-if-true}\tagsc{r-isZero-zero}\\[-5pt]
	\key{isZero} (\s \app e) & \step  \false     	\label{r-if-true}\tagsc{r-isZero-succ}\\[-5pt]
	(\lambda x. e)\app v & \step  e[v/x]    	\label{beta}\tagsc{beta}\\[-5pt]
	\head \app \nil & \step  \myRaise \app \zero    \label{r-head-nil}\tagsc{r-head-nil}\\[-5pt]
	{\head \app({{\cons \app v_1\app v_2}})} & \step {v_1}  
                                                        \label{r-head-cons}\tagsc{r-head-cons}\\[-5pt]
	\tail\app\nil & \step\myRaise\app(\s\app\zero)  \label{r-tail-nil}\tagsc{r-tail-nil}\\[-5pt]
	\tail \app(\cons \app v_1\app v_2) & \step v_2   \label{r-tail-cons}\tagsc{r-tail-cons}\\[-5pt]
	\key{isNil} \app (\nil) & \step \true	\label{r-head-nil}\tagsc{r-isNil-nil}\\[-5pt]
	\key{isNil}  \app(\cons  \app v_1\app v_2) & \step \false   	\label{r-head-nil}\tagsc{r-isNil-cons}\\[-5pt]
  \caseAbbr{(\key{inl}\app v)}{e_2}{e_3} & \step e_2[v/x_1]\label{r-inl}\tagsc{r-case-inl}\\[-5pt]
  \caseAbbr{(\key{inr}\app v)}{e_2}{e_3} & \step e_3[v/x_1]\label{r-inr}\tagsc{r-case-inr}\\[-5pt]
	\unfold \app (\fold \app v) & \step v     	\label{r-unfold-fold}\tagsc{r-unfold-fold}\\[-5pt]
	\fix \app v & \step v\app(\fix \app v)   	\label{r-fix}\tagsc{r-fix}\\[-5pt]
	\letrec{x}{v}{e} & \step e[(\fix\app (\lambda x.v))/x] 	\label{r-letRec}\tagsc{r-letRec}\\[-5pt]
	\try{v}{e}& \step v    	                        \label{r-try-ok}\tagsc{r-try}\\[-5pt]
	\try{(\myRaise \app v)}{e}& \step (e\app v)     \label{r-try-handler}\tagsc{r-try-raise}
\end{align*}
$$
\inference
	{e\step e'}
	{E[e] \step E[e']}\,\,\textsc{(ctx)}
\qquad\qquad 
F[\mathit{er}] \step \mathit{er}\,\,\textsc{(err-ctx)}
$$
\caption{The static and dynamic semantics of $\lambdafull$.}
\label{fig:LambdaFull:semantics}
\end{figure}

We repeat the statement of type soundness.  As usual, $\step^{*}$ is the
reflexive and transitive closure of $\step$.

\begin{center}
\shadowbox{\parbox{7.50cm}{\it
~\\[0.2ex]
\centerline{\large \textsc{Type Soundness Theorem:}}
for all expressions $e$, $e'$, 
and types $T$,\\
  if $\app\emptyset \typeOf \app  e: T$ and $e\step^{*}e'$ then either
\begin{itemize}
\item $e'$ is a value, 
\item $e'$ is an error, or
\item there exists $e''$ such that $e'\step e''$.
\end{itemize}
}}
\end{center}

Intuitively, when programs are well-typed they end up in a value or an
error, or the computation is simply not finished and continues.  A
well-typed program does not get stuck in the middle of a computation, that is,
\emph{well-typed programs cannot go wrong} (Robin Milner \cite{Milner78}).

\section{A Classification of the Operators}
\label{sec:classification}

A definition of a typed language such as that of Figure \ref{fig:LambdaFull} 
does not make important distinctions between the role of operators. 
Indeed, \key{cons}, \key{unfold} and \key{try} are grouped together 
within the same syntactic category Expressions, 
even though they play a very different role within the language. 
Operators can be classified in \emph{constructors},
\emph{eliminators}, \emph{derived operators}, and \emph{error
  handlers}. 
  
 In this section, we show a method for 
classifying operators into these classes. 
This method will be employed in Section \ref{sec:typeSystemForTypedLanguages} to automatically 
classify operators for language specifications given as input. 

\paragraph{Constructors}
Some operators of the language build values of a certain
type.  Those operators are called \emph{constructors}. 
We recognize them by the following characteristics.
\begin{quote}
\it Constructors have a typing rule whose assigned type is a constructed
type. Each constructor builds one value and each value is built by a constructor. 
Also, constructors have no reduction rules. 
\end{quote}
In $\lambdafull$, $\true$ and $\false$ are constructors for the type
$\Bool$. $\lambda x.e$ is constructor for the type $\rightarrow$, and $\nil$ and
$\cons\app e\app e$ are constructors for the type $\List$, 
to name a few examples.

\paragraph{Eliminators}
Eliminators can manipulate values of some type. 
For example, $\head \app e$ 
extracts the first element of the list $e$ when $e$ is reduced to a value. 
Some other operators simply inspect the identity of a value
such as $\key{if}$ operator. 
Eliminators have the following characteristics.
\begin{quote}
\it The typing rule of eliminators assigns a constructed type to one
 of their arguments: this argument is called the
\emph{eliminated argument}.  In all the reduction rules for
eliminators, the eliminated argument is pattern-matched against a
value. For convenience, we say that the rule \emph{eliminates that argument}.
\end{quote}
For example, the eliminated argument of $\key{if}$ is the first and
we say that \textsc{(r-if-true)} eliminates the first argument. 
%


\paragraph{Derived Operators}
Some operators are not involved in manipulating 
values at a primitive level. This is the case of
operators such as $\fix$ and $\letrecExp$, for example. These
operators are called \emph{derived operators}.
Derived operators have the following characteristics. 
\begin{quote}
\it Derived operators have at least one reduction rule. 
Also, none of their reduction rules pattern-matches
against a constructed expression. 
\end{quote}

\paragraph{Error Handlers}
It is often useful to capture an error produced by a computation
and trigger some remedial action. To this end, programming
languages with errors are sometimes augmented with operators
that can recognize the occurrence of errors and act accordingly. 
These latter operators are \emph{error handlers}. 
One of the most notable examples in programming languages, also
present in $\lambdafull$, is \key{try}.
Error handlers have the following characteristics. 
\begin{quote}
\it Error handlers have at least one reduction rule in which one 
of its arguments pattern-matches against an error. 
Analogously to eliminators, we call this argument the \emph{eliminated argument}. 
\end{quote}

\paragraph{Common Patterns}
Outside of the classification of operators, languages typically follow some common patterns 
for the sake of good design and type soundness. 

A value definition such as $\text{Values}   ::=  \ldots \mid
\cons \app v\app v$ tells us that the operator \key{cons} can build a
value only under some condition: that its two arguments are
evaluated to values.  These are \emph{valuehood requirements} that 
dictate when the definition can be applied. 
Valuehood requirements are used in error definitions (see $\text{Errors}
 ::=  \key{raise}\app v$), context definitions 
(for example, $\text{Contexts} ::=   v\app E$) and also for firing reduction
rules (see $\fix \app v \step v\app(\fix \app v)$). 
We adopt the following 
\begin{quote}
  \it \textbf{P-Val:} Value, error, and context definitions, as well as the firing of reduction rules can 
  depend only on valuehood requirements. 
\end{quote}

Also, languages typically conform to the following restrictions

\begin{quote}
  \it \textbf{P-NoStep:} Values and errors do not have reduction rules.
\end{quote}
\begin{quote}
\it \textbf{P-Typ:} Each operator has one typing rule and this typing rule assigns a type to 
each argument of the operator. 
\end{quote}

\section{A Discipline for the Progress Theorem}
\label{sec:progress:methodology}

In this section we spell out a methodology 
for ensuring the validity of the progress theorem.  
We first repeat its statement below. 

\begin{center}
\shadowbox{\parbox{7.50cm}{\it
~\\[0.2ex]
\noindent An expression $e$ \emph{progresses} whenever either
$e$ is a value, $e$ is an error, or there exists $e'$ such that $e\step e'$.
\medskip

\centerline{\large \textsc{Progress Theorem:}}
For all expressions $e$ and types $T$,\\
\centerline{if $\emptyset\typeOf \app  e: T$ then $e$ progresses.}
}}
\end{center}

We list the items of the methodology below as a convenient reference. 
Each item, except for \textbf{D0} which has been addressed, is described 
in detail in the following subsections.

\begin{enumerate}
\item[\textbf{D0}] Classify the operators of the language in \emph{constructors},
\emph{eliminators}, \emph{derived operators}, \emph{error
  handlers} and follow the common patterns as described in Section \ref{sec:classification}.
\item[\textbf{D1}] Progress-dependent arguments are contextual 
(this type of arguments is defined in Section \ref{sec:progress-arguments}).
\item[\textbf{D2}] Error contexts are evaluation contexts minus the error handler at the
  eliminated argument. 
\item[\textbf{D3}] The context declarations have no circular dependencies.
\item[\textbf{D4}] Each eliminator of a type eliminates all the values of that
  type.
  \item[\textbf{D5}] Error handlers have a reduction rule that is 
defined for values at their eliminated argument. 
  
\end{enumerate}

\subsection{D1. Progress-dependent Arguments}\label{sec:progress-arguments}
Consider the following definitions and reduction rules. 
\begin{align*}
   \text{Values}  &  ::=  \cons \app v\app v \mid \fold\app v  \\
   \text{Errors} & ::=  \key{raise}\app v  \\
  \text{Contexts} & ::=   v\app E\\
	\fix \app v & \step v\app(\fix \app v) 	\tagsc{r-fix}\\
	(\lambda x. e)\app v & \step  e[v/x]    \tagsc{beta}\\
	\try{(\myRaise \app v)}{e}& \step (v \app e)
                                        \tagsc{r-try-raise}
\end{align*}
In all the cases above some
arguments are under the restriction to be values or the error (an error is pattern-matched by \textsc{(r-try-raise)}). 
This is true also for \textsc{(beta)} w.r.t. the eliminated argument, 
where a value is syntactically pattern-matched.

These arguments need to be evaluated so that they become a value or error to 
enable the definition or reduction rule to apply. 
Therefore, they need to be in evaluation contexts. 
For example, since the argument of \key{fix} is required to be a value for \textsc{(r-fix)} to fire, 
$\lambdafull$ automatically needs to have the context Context ::= $\fix \app E$. 
Were the language to miss such context, the expression $\fix \app (\key{head}\app (\key{cons}\app  \lambda x.x \app\app \key{nil}))$, which is not a value nor an error, would be stuck.

We call the arguments that need to become values or errors \emph{progress-dependent
  arguments}. The way to identify them is the following.
\begin{quote}
\it Arguments that are required to be values in value, error and
contexts definitions are progress-dependent. \\
Arguments in the source of reduction rules that are required to be
values are progress-dependent. \\
Eliminated arguments are progress-dependent. 
\end{quote}


\begin{quote}
\textbf{D1} \it Evaluation contexts include all the progress-dependent arguments.
\end{quote}

Notice that \textbf{D1} leaves open the possibility of evaluation contexts for arguments that are not progress-dependent. 
Consider for example a $\lambda$-calculus with contexts Context ::= $(E\app e)\mid (e\app E)$, that is, the application evaluates its two arguments in parallel. Also consider the reduction rule $\beta' = (\lambda x.e_1)\app e_2 \step e_1[e_2/x]$. 
The first argument is certainly progress-dependent while the second, not encountering any restriction, is not. 
In this case, whether the second argument is contextual or not does not affect type soundness because a reduction happens either way thanks to $\beta'$ or a contextual step of the first argument. 



\subsection{D2. Error Contexts}

Language designers define the error contexts. However, not every error
context is suitable. The following is a general rule. 

\begin{quote}
\textbf{D2} \it Error contexts are evaluation contexts 
minus the error handler at the eliminated argument. 
\end{quote}

Error contexts do not contain the error handler at the eliminated argument for 
as a design choice. Indeed, $\try{(\myRaise \app e_1)}{e_2} \step \myRaise \app
e_1$ should not take place, as we expect the semantics of $\tryExp$ to
handle the error.

All other evaluation contexts are error contexts because the error handler is the only operator expecting an error. 
Therefore, all other progress-dependent arguments 
expect a value (by \textbf{P-Val}) and they have no reduction rule for handling the encounter of the error. 
For example, $\key{succ}\app (\myRaise \app v)$ would be stuck if it were not for the error context\footnote{We recall that Figure \ref{fig:LambdaFull} uses the meta variable $F$ for error contexts.} $(\key{succ} \app F)$ that 
enables the reduction $\key{succ}\app (\myRaise \app v)\step (\myRaise \app v)$. 

Strictly speaking, evaluation contexts for arguments that not progress dependent do not need to be in error contexts. 
For example, for the parallel $\lambda$-calculus of the previous section the expression $e\app (\myRaise \app v)$ does not get stuck because another reduction rule fires anyway. However, the language designer chooses evaluation contexts as such because those are the \emph{observable} parts of the expression, hence \textbf{D2} is generally the rule at play. 

For the same reason, \emph{only} evaluation contexts should be error contexts.   
For example, the error context $\ifExp{e}{e}{F}$ disregard the evaluation contexts of \key{if}, 
and allows the reduction $\ifExp{\true}{\zero}{{(\myRaise \app v)}}\step (\myRaise \app v)$. Of course, this reduction should not take place. 


\subsection{D3. Context Declarations}\label{circular}

\paragraph{A Problem with Dependencies} 
Consider the bad context declarations Context ::= $\key{cons}\app E\app
v\mid\key{cons}\app v\app E$. In this case, the expression 
$\key{cons}\app ((\lambda x.x) \app 5)  \app ((\lambda x.x) \app\nil)$ 
is simply stuck because the first argument $((\lambda x.x) \app 5)$ waits for the second to be reduced to a value, and at the same tiime $((\lambda x.x) \app\nil)$ waits for the first argument to be reduced to a value. 
Circular dependencies in context declarations jeopardize the 
type soundness of the language. Therefore, 

\begin{quote}
\textbf{D3} \it  Evaluation contexts must not have circular dependencies.
\end{quote}

An easy way to check for \textbf{D3} is through a graph representation of the dependencies at play. 
To be precise, for each declaration we have an edge from the index position of $E$ to the index position of a $v$. 
$\lambdafull$ has correct context declarations for \key{cons} because those declarations induce the graph $\{2 \mapsto 1\}$, which is acyclic. The bad context declarations above induce the graph $\{1 \mapsto 2,2 \mapsto 1\}$, which contains a cycle. 
\textbf{D3} accommodates most, if not all, of the common evaluation strategies in programming languages, such as left-to-right evaluation, right-to-left evaluation and also parallel evaluations.

\subsection{D4. Eliminators}

\begin{quote}
\textbf{D4}  \it For each eliminator of a type $T$,
  each value of type $T$ is eliminated by a reduction
  rule of the eliminator. 
\end{quote}
As an example, let us consider the $\head$ operator.
\begin{align*}
	\head \app \nil  \step  \myRaise \app\zero    	\tagsc{r-head-nil}\\
	\head \app(\cons \app v_1\app v_2)  \step v_1   \tagsc{r-head-cons}
\end{align*}

%
%
Were we to miss the rule \textsc{(r-head-nil)}, the expression $(\head \app \nil)$ would be stuck for 
not finding a reduction rule that fires. 
As this expression is not a value nor an error, type soundness would be jeopardized.

\subsection{D5. Error Handlers}
\label{ssec:errors}

\begin{quote}
\textbf{D5}  \it  Error handlers have a reduction rule that is 
defined for values at their eliminated argument. 
\end{quote}

In $\lambdafull$, the error handler \key{try} cannot afford to define its step only at the encounter of the error, 
otherwise an expression such as $\try{\zero}{\lambda x.x}$ would be stuck. 
\textbf{D5} imposes that a reduction 
rule such as 
\begin{align*}
  \try{v}{e} \step v    	                \tagsc{r-try}
\end{align*}
exists. Notice that the rule expects precisely a value. 
Indeed, we should forbid rules of the form $\key{try}\app e\app \key{with} \app (\emph{some expression}) \step (\emph{some expression})$ that apply unrestricted on the eliminated argument. 
As the error is also an expression, 
this rule can non-deterministically preempt the application of the rule that specifically handles the error. 

\section{Type preservation}
\label{sec:typePreservation}

We now devise a methodology for checking the validity of 
the type preservation theorem. First, we repeat the statement of the theorem.
\begin{center}
\shadowbox{\parbox{7.50cm}{\it
~\\[0.2ex]
\centerline{\large \textsc{Type Preservation Theorem :}}\\
for all expressions $e$, $e'$
and types $T$, \\
\centerline{if $\app \emptyset \typeOf \app
  e: T$ and $e \step e' $ then $\emptyset \typeOf \app
  e': T$}\\
}}
\end{center}
Given a reduction rule $e \step e' $, we have to ensure that the types
of $e$ and $e'$ coincide. However, this rule makes use of 
variables that can be instantiated to a plurality of expressions.  
Ideally, we need to check that 
\[
\text{for all } \Gamma,~ \Gamma \typeOf e:T \text{ implies } \Gamma \typeOf e':T.
\]
Of course, checking all possible type environments is
prohibitive. Therefore, our approach approximates such a check with the
use of a \emph{symbolic type environment}. 
We form symbolic type environments out of the typing rules of operators. 
For convenience, we simply use the typing premises
that we encounter in those rules. 
This choice accommodates well the fact that typing premises rely on 
typing assumptions themselves. 
Consider for example the premise  
$\Gamma,x:T_1 \typeOf \app e : T_2$ of \textsc{(t-abs)} 
 and $\textit{exp} =  \lambda x.v$. Variables
 have two levels. 
Typing \textit{exp} depends on $v$, which is the logical variable of the typing rule and 
ranges over expressions. In turn, after $v$ is instantiated, it contains a particular variable 
$x$ of the object language, and the type of $v$ depends on this variable. 
To account for this, the symbolic type environment employs hypothetical typing formulae. 
For example, the symbolic type environment extracted for $\textit{exp}$ is 
$(\Gamma,x:T_1 \typeOf \app v : T_2)$. The presence of hypothetical typing formulae is axiomatized by
the following equation.
\begin{gather*}
\inference{\Gamma \typeOf e':T_1}{\Gamma,x:T_1 \typeOf \app e : T_2
  \equiv\app \Gamma \typeOf e[e'/x] : T_2}\tagsc{eq-sub}
\end{gather*}

Given a reduction rule, we give a means to compute both the symbolic type environment 
and its symbolic assigned type. 
There are two steps for
those reduction rules that eliminate an argument ((1) and (2) below) and 
one step for any other reduction rule (only (1)).
\begin{quote}
\it (1) Instantiate the typing rule that types the source of
the reduction rule. 
(2) Instantiate the typing rule that types the eliminated 
argument of the reduction rule, if that argument is a constructed expression. 
The symbolic type environment contains the typing formulae 
of the premises of the 
two rules combined. 
The symbolic assigned type is that of (1). 
\end{quote}

With this main ingredient, we can offer a methodology for
type preservation. 
For each reduction rule, apply the following.

\begin{quote}
\it Construct the symbolic type environment $\Gamma^{s}$ of the rule 
and its symbolic assigned type $T$. 
Check whether $\Gamma^{s}$ entails that
the target of the reduction rule has the same type $T$. 
\end{quote}

We shall see a few examples. Consider the case of $\head$
and its elimination rule $\head \app(\cons \app v_1\app v_2)  \step \blue{v_1}$.
We have given the color blue to the target so that later 
it will be clear where a particular occurrence of $v_1$ comes from. 
Instantiating the typing rules \textsc{(t-head)} and \textsc{(t-cons)} in
the way prescribed by (1) and (2), respectively, gives us the following rules. 

\begin{gather*}
\inference
	{\Gamma \typeOf \app (\cons \app v_1\app v_2) : \List \app T} 
	{ \Gamma \typeOf \app \head \app (\cons \app v_1\app v_2) :
          \redd{T}}
\, 
\inference
	{
	\Gamma \typeOf v_1 : T \\
	\Gamma \typeOf v_2 : \List\app T
	 }
	{ \Gamma \typeOf \app \cons \app v_1\app v_2:  \List \app T}
\end{gather*}


The assigned type is the red $\redd{T}$
in the first rule. For the symbolic type assignment, we collect the
typing premises of the two rules. 
We can restrict ourselves to collect only the typing rules for variables. 
Indeed, the typing premise of the eliminated argument, such as $\Gamma \typeOf \app (\cons \app v_1\app v_2) : \List \app T$, is always derivable because it has been unfolded in the second rule. 
For the case above, we have $\Gamma^{s} = v_1 : T, \,v_2 : \List\app T$. 
Finally, we need to check
that $\Gamma^{s} \vdash \blue{v_1} : \redd{T}$. This fact can be trivially established. 
This means that
$\Gamma^{s}$, which can type $\head \app(\cons \app v_1\app v_2)$ at $\redd{T}$, can also type $\blue{v_1}$ at $\redd{T}$. 



Let us now see the example of \textsc{(beta)}: $(\lambda x. e)\app v
\step  \blue{e[v/x]}$. 
The instantiations (1) and (2) give us the following rules.
\begin{gather*}
\inference
	{
	\Gamma \typeOf \app  \lambda x. e : T_1\to T_2 \\
	\Gamma \typeOf \app v : T_1 	
	} 
	{ \Gamma \typeOf \app  ((\lambda x. e)\app v) : \redd{T_2}}
\quad
\inference
	{\Gamma,x:T_1 \typeOf \app e : T_2} 
	{ \Gamma \typeOf \app \lambda x. e:  T_1\to T_2}
\end{gather*}
In this case, the symbolic type environment is 
$(\Gamma,x:T_1 \typeOf \app e : T_2),\, \Gamma \typeOf v :
T_1$. We finally need to check  
$\Gamma^{s} \vdash \blue{e[v/x]} : \redd{T_2}$, which can be
established using \textsc{(eq-sub)}.

\paragraph{A Requirement for Errors}
In a language with errors and error contexts, we enforce that 

\begin{quote}
\textbf{D-Err} \it Errors must be typed at any type.
\end{quote}

This is necessary because errors travels through contexts thanks to the rule $F[\mathit{er}] \step \mathit{er}$, 
for any context $F$. For type preservation, wherever an error lands it must be prepared to match 
the type of the expression it replaces. 

\paragraph{Some Remarks}

In the next sections we strive to model the methodologies of this section of Section \ref{sec:progress:methodology} as type systems. 
The methodology for progress is markedly a discipline and as such 
it can be easily seen as a type system. 
On the other side, the methodology for type preservation does not leave much room for a discipline. 
Language designers in the first place do not employ a particular discipline but they simply write reduction rules 
according to the meaning of their operators. 
Nonetheless, a formulation of this methodology as a type system is as natural.
The analogy is with programs. Type systems for programs typically have parts where a precise discipline is enforced 
and parts that merely perform checks. Consider the \key{if} operator of $\lambdafull$ and its typing rule. 
\[
\inference
	{\Gamma \typeOf \app e_1 : \Bool &
	\Gamma \typeOf \app e_2 : T &
	\Gamma \typeOf \app e_3 : T 
	} 
	{ \Gamma \typeOf \app\ifExp{e_1}{e_2}{e_3} : T}  \,\,\textsc{(t-if)}
\]
and let us focus on $e_2$ and $e_3$. 
While many other parts of the type system of $\lambdafull$ enforce a precise discipline, 
this particular part simply checks that $e_2$ and $e_3$ have the same type with a same type environment. 
In many ways, this is the discipline imposed. 
Similarly, and equally naturally, when we type check a reduction rule $e_1 \step e_2$ we simply check that
$e_1$ and $e_2$ have the same type for same type environments, and this is the discipline imposed.

So far, we have spelled out a descriptive methodology for ensuring both the progress 
and preservation theorem. It is easy to check that it applies to $\lambdafull$ in full. 
This is a non-trivial language with modern features such as recursive types, polymorphism and exceptions. 
Now that we have described the methodology in detail we can proceed to formalize it as a typing discipline 
and prove it correct. 
\section{Typed Languages as Logic Programs}
\label{sec:typeLanguagesAsLogicPrograms}

We now proceed to give the methodologies of the previous sections a formal counterpart.  
To this aim, we first need a formal representation for language specifications that 
can be manipulated and be the subject of proofs. 
We represent them as logic programs in the higher-order intuitionistic logic. 
This logic has a solid theoretical foundation, is executable and is the basis of the $\lambda$Prolog programming language. 
Higher-order logic programs turn out to be a convenient medium for our endeavors also because they are 
in close correspondence to pen\&paper language definitions. 

Logic programs are equipped with a \emph{signature}, which is a set of 
\emph{declarations} for the entities that are
involved in the specifications.  For example, the following is a
partial signature for $\lambdafull$. 
\[
\small{
\begin{array}{l}
\lp{exp}, \lp{type} : \lp{kind} \\
\lp{arrow} : \lp{type} \to \lp{type} \to \lp{type}\\
\lp{abs} : \lp{type} \to (\lp{exp} \to \lp{exp}) \to \lp{exp} \\
\lp{app} : \lp{exp} \to \lp{exp} \to \lp{exp}\\
\end{array} 
}
\]

\textit{Convention:} Some parts of the meta type system in Section \ref{sec:typeSystemForTypedLanguages} and some parts of the language typing rules being typed may look very similar. To avoid confusion, we display parts of logic programs in \blue{blue color} and, additionally, constants are displayed in $\small{\lp{bold sans-serif font}}$. 

 The schema variable $\Sigma$ ranges over 
signatures. A specific constant, precisely the constant $\small{\proptype}$, is the type of propositions. 
To help the presentation, we sometimes use symbols rather than names. 
For example, we adopt the declarations 
$\small{\typeofTL : \lp{exp} \to \lp{type} \to \proptype}$ and 
$\small{\stepTL : \lp{exp} \to \lp{exp} \to \proptype}$ for a typing and a reduction predicate, respectively. 
We shall use $\stepTL$ in infix notation and write 
$\small{\blue{e_1\app\stepTL \app e_2}}$ rather than $\small{\blue{\stepTL\app e_1 \app e_2}}$. 
Also, we shall keep using the familiar ``:" in typing formulas as a slight abuse of notation. 
For example, we write $\small{\blue{\typeofTL\app e : T}}$ rather than $\small{\blue{\typeofTL\app e \app T}}$.
Given a signature $\Sigma$, we denote by $\Sigma(\lp{\small{exp}})$ and $\Sigma(\lp{\small{type}})$ the sets 
of constants in $\Sigma$ that define expressions and types, respectively. 
In \lambdafull, $\Sigma(\lp{\small{exp}})$ contains \lp{\small{abs}}, \lp{\small{app}}, \lp{\small{head}}, \lp{\small{tail}}, \lp{\small{cons}}, \lp{\small{nil}}, \lp{\small{fold}}, and \lp{\small{unfold}}, among the rest of operators. 
Likewise, $\Sigma(\lp{\small{type}})$ contains \lp{\small{arrow}}, \lp{\small{bool}}, \lp{\small{int}}, \lp{\small{list}}, \lp{\small{forall}}, \lp{\small{mu}} (recursive type), and \lp{\small{sum}}.

When we represent program expressions as types, we shall
use the familiar setting of higher order abstract syntax (HOAS) to
encode bindings.  That is, binders in program expressions will be
mapped directly to binders in terms.  For example, the declaration of
the abstraction \lp{\small{abs}} above takes two parameters, of which the second is
an abstraction of the logic.  The identity function $\lam{x\of \Bool}
x$ is then encoded as $\small{(\lp{abs}\,\, \lp{bool} \,\, \blue{\lam{x} x})}$.

The \emph{terms} of higher-order logic are based on the usual notion
of simply typed $\lambda$-terms over a signature. We use the symbol $t$ to range over terms. 
Given a signature $\Sigma$, a (higher-order intuitionistic logic)
formula $\small{\blue{P}}$ over $\Sigma$ is any formula built from implications and
universal quantifier and atomic formulas. 
We shall represent logic programming rules $\phi$ in the form  
$$
\blue{
\inference{
P_1 \app \ldots \app P_n
}{P}}
$$

%

In higher-order logic programs, the use of universal and implicational formulas 
enable generic and hypothetical reasoning. 
Their role in language specification can be described with the example of the following typing rule 
for the abstraction operator $\small{\blue{\lp{abs}}}$. 
\[
\small{
\blue{
\inference{
  (\forall x.\app \typeofTL\app  x : T_1 \Rightarrow\, \typeofTL \app (E ~x): T_2)
}{
  \typeofTL \app (\ensuremath{\lp{abs}} ~T_1~E):
    (\ensuremath{\lp{arrow}} ~ T_1 ~T_2)
}
}
}
\]
The universal quantification $\forall x$ introduces a new variable $x$ 
encoding a program term and the implication temporarily augment 
the logic program with the fact $\small{\blue{\typeofTL \app x : T_1}}$ while proving 
$\small{\blue{\typeofTL \app (E ~x): T_2}}$. 
Therefore, the explicit type environment $\Gamma$, which encodes the typing information 
assumed along the way, is not necessary. 
%
%

\newcommand{\N}{\mathbb{N}}
\newcommand{\Power}[1]{\mathcal{P}(#1)}


Our notion of typed languages is based on the following standard definition of logic programs. 

\begin{definition}[Logic Programs]\label{def:LogicPrograms}
A \emph{logic program} is a pair $(\Sigma, D)$ where $\Sigma$
is a signature and $D$ is a set of rules over $\Sigma$. 
A query $q$ (which can be any logical formula) follows from a logic
program, written $(\Sigma, D) \models \small{\blue{P}}$, if $\small{\blue{P}}$ is provable from $D$
in intuitionistic logic.
\end{definition}

As we have seen in the previous sections, typed languages also rely on
evaluation and error contexts. We define \emph{context summaries} as a 
declarative means for their specification. Intuitively, the following contexts 
$\key{head}\app E \mid \key{raise}\app E \mid E\app e \mid v\app E 
\mid \key{cons}\app E\app e \mid \key{cons}\app v\app E$ from the $\lambdafull$ language
are modeled with a function $\contextsTL$ such that 
\begin{align*}
\contextsTL(\blue{\small{\lp{head}}}) = & \contextsTL(\blue{\small{\lp{raise}}}) = \{(1,
\emptyset)\}\\
\contextsTL(\small{\lp{app}}) = & \contextsTL(\small{\lp{cons}}) = \{(1, \emptyset), (2, \{1\})\}.
\end{align*}

Here, $(2, \{1\})$ means that the second argument is contextual but
requires the first to be a value. 
\begin{definition}[Context summaries]\label{def:contextSummary}
Given a signature $\Sigma$, a \emph{context summary over $\Sigma$} is a function 
$\contextsTL$ from $\Sigma(\lp{\small{exp}})$ to $\Power{\N \times \Power{{\N}}}$.
\end{definition}

For an operator $\small{\blue{\lp{op}}}$, we simply write 
$\{i_1, \ldots i_n\} \subseteq \contextsTL(\small{\lp{op}})$ to mean $\{i_1, \ldots i_n\} \subseteq \texttt{dom}(\contextsTL(\small{\lp{op}}))$. For example, given the definitions above we have 
$\{1,2\}\subseteq \contextsTL(\small{\lp{cons}})$.

Typed languages are logic 
programs augmented with two context summaries for evaluation and error contexts. 

\newcommand{\errContextsTL}{\textit{err\textendash ctx}}

\begin{definition}[Typed Languages]\label{def:TypedLanguages}
A \emph{typed language} is a tuple $(\Sigma, D, \contextsTL,
\errContextsTL)$ such
that
\begin{itemize}
\item ($\Sigma$, D) is a logic program, such that $\Sigma$ contains
  kinds $\small{\lp{exp}}$ and $\small{\lp{type}}$ and 
\[
\small{
\begin{array}{l}
\typeofTL : \lp{exp} \to \lp{type} \to \proptype.\\
\stepTL : \lp{exp} \to \lp{exp} \to \proptype . \\
\multistepTL : \lp{exp} \to \lp{exp} \to \proptype . \\
\valueTL : \lp{exp} \to \proptype. \\
\errorTL : \lp{exp} \to \proptype. 
\end{array} 
}
\]
\item $\contextsTL$ is a context summary over $\Sigma$. 
\item $\errContextsTL$ is either $\mathrm{None}$ or is a context summary over $\Sigma$. 
\item  $D$ contains the rules that define $\multistepTL$ as the 
reflexive and transitive closure of $\stepTL$.%
\end{itemize}
\end{definition}

We let $\Ldl$ range over typed languages. We use \blue{$E$} for variables of kind $\small{\lp{exp}}$, and \blue{$T$} for
those of kind $\small{\lp{type}}$. 
Terms of kind $\small{\lp{exp}}$ are ranged over by \blue{$e$} and those of kind $\small{\lp{type}}$ by 
\blue{$ty$}. 
We use the notation $D_{\mid \mathit{\blue{pred}}}$ to denote the subset of rules in $D$ that define the predicate $\mathit{\blue{\small{pred}}}$. For example, $D_{\mid \typeofTL}$ and $D_{\mid\stepTL}$ are the typing and the reduction rules in $D$, respectively. 
The semantics of a typed language $\Ldl$ is defined as its straightforward counterpart
as logic program in which context summaries are translated into rules. For
example, $\contextsTL(\small{\lp{cons}}) = \{(1, \emptyset), (2, \{1\})\}$
generates the two rules below. 
\begin{gather*}
\small{\blue{
\inference
{ E_1\app\stepTL \app E_1'}
{ (\lp{cons}\app  E_1\app E_2)\app\stepTL \app (\lp{cons}\app E_1'\app E_2)}
}}
\quad
\small{\blue{
\inference
{\valueTL\app E_1 & E_2\app \stepTL \app E_2'}
{ (\lp{cons}\app  E_1\app E_2)\app\stepTL \app (\lp{cons}\app E_1\app E_2')}
}}
\end{gather*}

\noindent and $\errContextsTL(\small{\lp{head}}) = \{(1, \emptyset)\}$ generates 
$\small{\blue{
\inference
{\errorTL \app E}
{(\lp{head}\app  E)\app\stepTL \app E}
}}
$.

We overload $\models$ to typed
languages, with the meaning that typed languages are first translated to
logic programs. 

\paragraph{Syntactic Sugar for Representing Languages} 
In the next section we develop meta type systems that inspect logic programming 
based representations of languages. 
To help our presentation, we employ some syntactic sugar to the raw syntax so far introduced to make it closer to familiar syntax in language design.

Typing rules are augmented with a type environment for replacing generic and hypothetical occurrences. The symbol for the type environment is fixed to be $\blue{\lpGamma}$. Below we show the typing rules \textsc{(t-tail)}, \textsc{(t-abs)}, and \textsc{(t-abst)} as logic programming rules on the left. They are an example on how typing rules from $\lambdafull$ are modeled in our context. On the right, we then show the counterpart syntax we adopt in the next section. 
\begin{gather*}
\small{
\blue{
\inference
	{ \typeOf \app e : \lp{list} \app T} 
	{  \typeOf \app \lp{tail} \app e:  \lp{list}\app T}
\equiv 
\inference
	{\lpGamma \typeOf \app e : \lp{list} \app T} 
	{\kern -5pt\lpGamma \typeOf \app \lp{tail} \app e:  \lp{list}\app T\kern -5pt}
	}}
\\[1.5ex]
\small{
\blue{
\inference{
  (\forall x.\app \typeofTL\app  x : T_1 \Rightarrow\, \typeofTL \app (E ~x): T_2)
}{
  \typeofTL \app (\ensuremath{\lp{abs}} ~T_1~E):
    (\ensuremath{\lp{arrow}} ~ T_1 ~T_2)
}
\equiv
\inference{\lpGamma, x:T_1 \app \typeofTL  \app E : T_2}
	{\lpGamma\app  \typeofTL \app(\ensuremath{\lp{abs}} ~T_1~E):
    (\ensuremath{\lp{arrow}} ~ T_1 ~T_2)}
}}
\\[1.5ex]
\small{
\blue{
\inference
{\forall x. \typeOf \app E : (T\app x)} 
          { \typeOf \app \lp{absT} \app E :  (\lp{forall} \app T)}
\equiv
\inference
{\lpGamma, x \typeOf \app E : T} 
          { \lpGamma\typeOf \app \lp{absT} \app E :  (\lp{forall} \app T)}
}}
\end{gather*}
We adopt the convention that variables $\small{\blue{V}}$ are treated as \emph{value variables} and entail that the rule implicitly contains the premise $\blue{\small{\valueTL\app V}}$. Below we model \textsc{(beta)} and \textsc{(r-head-cons)} on the left. These are examples of reduction rules of $\lambdafull$ modeled in our setting. On the right, we display how we represent them in the next section. 
\blue{
\begin{gather*}
\small{
\inference{\valueTL\app E_2}
{ (\lp{app}\app (\lp{abs} \app E_1) \app E_2) \app \stepTL \app(E_1\app  E_2)} 
\equiv 
 (\lp{app}\app (\lp{abs} \app E) \app V) \app \stepTL \app (E\app  V) 
}
\\[1.5ex]
\small{
\inference{\valueTL\app E_1 & \valueTL\app E_2}
{ \lp{head}\app (\lp{cons} \app E_1\app E_2) \app \stepTL \app E_1} 
\equiv 
 \lp{head}\app (\lp{cons} \app V_1\app V_2) \app \stepTL \app V_1
}
\end{gather*}
}
Value and error definitional rules are rewritten in the following style. Notice, below we have also applied the convention on value variables. 
\begin{gather*}
\small{\blue{
\inference{\valueTL\app E_1 & \valueTL\app E_2}
{\valueTL \app (\lp{cons} \app E_1\app E_2)} 
\quad \equiv \quad
\valueTL ::= (\lp{cons} \app V_1\app V_2) 
}}
\\ 
\small{\blue{
\inference{\valueTL\app E}
{\errorTL \app (\lp{raise} \app E)} 
\quad \equiv \quad
\errorTL ::= (\lp{raise} \app \app V) 
}}
\end{gather*}

Without loss of generality, we assume that type annotation arguments are always first. 
Also, we use a special notation for type-annotated operators. To make an example, we would display the type-annotated version of the operator $\small{\lp{cons}}$ with $(\small{\lp{cons}}[T] \app e_1\app e_2)$, and we use this notation throughout the paper. 

\begin{figure*}
{\scriptsize
\fbox{$ \vdash \Ldl $}
\begin{gather*}
\inference
	{
           (D_{|\text{$\valueTL$}} \cup D_{|\text{$\errorTL$}}) = \{\phi\defSup_1, \ldots, \phi\defSup_n\} &
           D_{|\text{$\typeofTL$}} = \{\phi\typSup_1, \ldots, \phi\typSup_m\}&
		D_{|\text{$\stepTL$}} = \{\phi\redSup_1, \ldots, \phi\redSup_l\} \\
           \contextsTL \typeOf\defSub \phi\defSup_1 : B\defSup_1 \app\ldots\app \contextsTL \typeOf\defSub \phi\defSup_n : B_n\defSup & \Gamma\defSup \equalUnique B\defSup_1, \ldots, B\defSup_n \\         
	D_{|\text{$\stepTL$}} \mid \Gamma\defSup \typeOf\typSub \app \phi\typSup_1 : B\typSup_1 \app 
		\ldots \app 
		D_{|\text{$\stepTL$}} \mid \Gamma\defSup  \typeOf\typSub \app \phi\typSup_m : B\typSup_m
		& \Gamma\typSup \equalUnique B\typSup_1, \ldots, B\typSup_m\\  
	 \contextsTL \mid  \Gamma\typSup \typeOf\redSub \phi\redSup_1 : B\redSup_1
	\app \ldots \app 
	\contextsTL \mid  \Gamma\typSup \typeOf\redSub \phi\redSup_l : B\redSup_l
	& \Gamma\typSup \exh (B\redSup_1, \ldots, B\redSup_m) \\
	& \texttt{well-formed}(\contextsTL)
& \contextsTL\mid \Gamma\typSup \typeOf  \errContextsTL\\
	\Ldl = (\Sigma, D, \contextsTL,\errContextsTL) &
	D_{|\text{$\typeofTL$}} \typeOf^{\Ldl}\preSub \phi\redSup_1
	\app \ldots \app 	
	D_{|\text{$\typeofTL$}} \typeOf^{\Ldl}\preSub \phi\redSup_l
		 	} 
	{ \typeOf  (\Sigma, D, \contextsTL,\errContextsTL)  }\quad\textsc{(ts-main)}
\end{gather*}  
\begin{gather*}
\texttt{well-formed}(\contextsTL)  \text{ iff } \begin{cases} 
\forall \blue{op}, 
\contextsTL(\blue{op})\emph{ is a directed acyclic graph}, \text{ and } \\
N\in \texttt{rng}(\contextsTL(\blue{op})) \text{ implies } N \subseteq \contextsTL(\blue{op})
 \end{cases} 
 \\[1.5ex]
\Gamma_1 \equalUnique \Gamma_2 \text{ iff } \begin{cases} 
\Gamma_1 = \Gamma_2, \text{ and }  \\
 \Gamma_1(\blue{op}) = B_1, \app \Gamma_1(\blue{op}) = B_2 \text{ implies } B_1 = B_2, \text{ and } \\
 \Gamma_1(\blue{op_1}) = \text{error} \app N_1, \app 
\Gamma_1(\blue{op_2}) = \text{error} \app N_2\text{ implies } \blue{op_1} = \blue{op_2} \text{ and } N_1 = N_2.
 \end{cases} 
 \\[1.5ex]
 \Gamma\typSup \exh \Gamma\redSup   \text{ iff } 
\begin{cases}  
\{ \blue{op_1} :  \text{elim} \app c, \app \blue{op_2} :\text{value} \app c\app N\} \subseteq \Gamma\typSup \text{ implies } \blue{op_1} :  \text{eliminates} \app \blue{op_2} \in \Gamma\redSup, \text{ and } \\
\{ \blue{op_1} :  \text{errorHandler}, \app \blue{op_2} :\text{error} \app N \} \subseteq\Gamma\typSup \text{ implies } 
\{\blue{op_1} :  \text{eliminates} \app \blue{op_2}, \blue{op_1} :  \text{plain} \} \subseteq \Gamma\redSup.
\end{cases} 
\end{gather*}  
\fbox{$   \contextsTL \typeOf\defSub \phi : B\defSup $} 
\begin{gather*}
\inference
	{ 	  
 \{1, \ldots,  n\}\subseteq \contextsTL(\blue{op})
 }
	{ \contextsTL \typeOf\defSub
	 \blue{\valueTL ::= (op[\widetilde{T}]\app V_1\app \cdots \app V_n \app\widetilde{E}) } : \blue{op} : \text{value} \app \{1,\ldots, n\} }\quad\textsc{(d-value)}
\\[1.5ex]
\inference
	{ 	  
 \{1, \ldots,  n\}\subseteq \contextsTL(\blue{op})
 }
	{\contextsTL \typeOf\defSub
	 \blue{\errorTL::= (op[\widetilde{T}]\app V_1\app \cdots \app V_n \app\widetilde{E}) } : \blue{op} : \text{error} \app \{1,\ldots, n\} }\quad\textsc{(d-error)}\\
\end{gather*}	 
 \fbox{$D\mid \Gamma\defSup \typeOf\typSub \app \phi : B\typSup $} 
\begin{gather*}
	D^{\text{r}}  \mid \Gamma\defSup, \blue{op} : \text{value}\app N \typeOf \blue{\inference{\lpGamma_1\app\typeofTL\app E_1 : ty_1& \ldots & \lpGamma_n\app\typeofTL\app E_n : ty_n}{\lpGamma\app\typeofTL\app (op[\widetilde{T_1}]\app E_1\app\cdots\app E_n):
          (c \app \widetilde{T_2})}} : \blue{op} : \text{value}\app \blue{c} \app N\quad\textsc{(t-value)}
\\[2ex]
\inference
	{
            \blue{(op_1 [\widetilde{T_2}]\app (op_2[\widetilde{T_3}]\app \widetilde{E\valueSup_1}) \app \widetilde{E\valueSup_2}) \app \stepTL\app e }\in D^{\text{r}}.\field{rules}(\blue{op_1})\\
            \Gamma(op_2) = \text{value}\app N
	} 
	{D^{\text{r}}  \mid \Gamma\defSup \typeOf \blue{\inference{\lpGamma_1\app\typeofTL\app E_1 : (c \app \widetilde{T})& \ldots & \lpGamma_n\app\typeofTL\app E_n : ty_n}
	{\lpGamma\app\typeofTL\app (op_1 [\widetilde{T_1}]\app E_1\app\cdots\app E_n):
          ty}} : \blue{op_1} : \text{elim}\app \blue{c} }\quad\textsc{(t-elim)}
\\[2ex]
\inference
	{ 
            \blue{(op_1 [\widetilde{T_2}]\app (op_2[\widetilde{T_3}]\app \widetilde{E\valueSup_1}) \app \widetilde{E\valueSup_2}) \app \stepTL\app e }\in D^{\text{r}}.\field{rules}(\blue{op_1})\\
            \Gamma(op_2) = \text{error}\app N
	} 
	{D^{\text{r}}\mid \Gamma\defSup \typeOf \blue{\inference{\lpGamma_1\app\typeofTL\app E_1 : ty_1& \ldots & \lpGamma_n\app\typeofTL\app E_n : ty_n}
	{\lpGamma\app\typeofTL\app (op_1 [\widetilde{T_1}]\app E_1\app\cdots\app E_n):
          ty}} : \blue{op_1} :  \text{errHandler}}\quad\textsc{(t-errHandler)}
\\[2ex]
\inference
	{ D^{\text{r}}.\field{rules}(\blue{op})  = \{\phi_1, \ldots, \phi_m\} & m \geq 1 \\
	\forall i,  1 \leq i \leq m, & \phi_i = \blue{(op [\widetilde{T_i'}]\app \widetilde{E\valueSup_i}) 
	 \app \stepTL\app e_i}
	} 
	{D^{\text{r}} \mid \Gamma\defSup \typeOf \blue{\inference{\lpGamma_1\app\typeofTL\app E_1 : ty_1& \ldots & \lpGamma_n\app\typeofTL\app E_n : ty_n}
	{\lpGamma\app\typeofTL\app (op [\widetilde{T}]\app E_1\app\cdots\app E_n):
          ty}} : \blue{op} : \text{derived}}\quad\textsc{(t-derived)}
\\[2ex]
\inference
	{
	\blue{T} \not\in \key{vars}(\blue{ty_1}) \cup \ldots \cup \key{vars}(\blue{ty_n}) 
		} 
	{D^{\text{r}} \mid \Gamma\defSup, \blue{op}: \text{error}\app N \typeOf \blue{\inference{\lpGamma_1\app\typeofTL\app E_1 : ty_1& \ldots & \lpGamma_n\app\typeofTL\app E_n : ty_n}
	{\lpGamma\app\typeofTL\app (op [\widetilde{T}] \app E_1\app\cdots\app E_n):
          T}} : \blue{op}: \text{error}\app N}\quad\textsc{(t-error)}
\end{gather*}  
}
  \caption{Type system for type soundness: main typing judgement, value and error definitions and typing rules. }
  \label{fig:typeSystemTL}
\end{figure*}

\begin{figure*}
{\scriptsize
\fbox{$   \contextsTL\mid \Gamma\typSup \typeOf  \errContextsTL$} 
\begin{gather*}
\inference
	{\textsc{(err-none)} & $\,\,$ & $\,\,$ \\
	\forall \blue{op}, \,\,\blue{op}:\text{error} \not\in\Gamma\typSup}
	{ \contextsTL\mid \Gamma\typSup \typeOf  \text{None} }
\quad 
\inference
	{\textsc{(err-only)} & $\,\,$ & $\,\,$& $\,\,$ & $\,\,$\\
	\forall \blue{op_2}, \,\,\blue{op_2}:\text{errHandler} \not\in\Gamma\typSup}
	{ \contextsTL \mid \Gamma\typSup, \blue{op_1}:\text{error} \typeOf \contextsTL }
\quad 
\inference
	{\textsc{(err-handler)} & $\,\,$ & $\,\,$& $\,\,$ \\
	\blue{op_1}:\text{error} \in\Gamma\typSup \\
	\blue{op_2}:\text{errHandler} \in\Gamma\typSup
	}
	{ \contextsTL \mid \Gamma\typSup \typeOf \contextsTL \setminus \{\blue{op_2} \mapsto (1, \emptyset)\} }
\end{gather*}  
\fbox{$\contextsTL \mid \Gamma\typSup \typeOf\redSub \phi:  B\redSup $} 
\begin{gather*}
\inference
	{ 	  
	\Gamma\typSup(\blue{op_1}) = \text{elim}\app \blue{c} & \Gamma\typSup(\blue{op_2}) = \text{value}\app \blue{c}\app N\\
	\widetilde{V}  = V_1  \app\cdots \app  V_n & N = \{1, \ldots, n\}\\
	\widetilde{V'}  = V'_2  \app\cdots \app  V'_m & \{1, \ldots, m\} \subseteq \contextsTL(\blue{op_1}) \\
           	} 
	{ \contextsTL \mid \Gamma\typSup \typeOf \blue{(op_1 [\widetilde{T_1}] \app
           (op_2 [\widetilde{T_2}] \app \widetilde{V} \app \widetilde{E}) \app \app\widetilde{V'} \app \widetilde{E'})\app\stepTL\app e}: \blue{op_1} : \text{eliminates}\app \blue{op_2}
           }\quad\textsc{(r-elim)}
\\[1.5ex]
\inference
	{ 	  
	\Gamma\typSup(\blue{op_1}) = \text{errHandler} & \Gamma\typSup(\blue{op_2}) = \text{error}\app N\\
	\widetilde{V}  = V_1  \app\cdots \app  V_n & N = \{1, \ldots, n\}\\
	\widetilde{V'}  = V'_2  \app\cdots \app  V'_m & \{1, \ldots, m\} \subseteq \contextsTL(\blue{op_1}) \\
           	} 
	{ \contextsTL \mid \Gamma\typSup \typeOf \blue{(op_1 [ \widetilde{T_1}] \app
           (op_2 [ \widetilde{T_2}] \app \widetilde{V} \app \widetilde{E}) \app \app\widetilde{V'} \app \widetilde{E'})\app\stepTL\app e} : \blue{op_1} : \text{eliminates}\app \blue{op_2}}\quad\textsc{(r-errHandler)}
\\[1.5ex]
\inference
	{ 	  
	\widetilde{V}  = V_2  \app\cdots \app  V_n & 
	 \{1, \ldots, n\}\subseteq \contextsTL(\blue{op})\\
	     	} 
	{ \contextsTL \mid \Gamma\typSup, \blue{op} : \text{errHandler}  \typeOf \blue{
           (op [\widetilde{T}] \app V_1\app \widetilde{V}\app \widetilde{E})\app\stepTL\app e } : \blue{op}: \text{plain}}\quad\textsc{(r-errHandler-value)}
\\[1.5ex]
\inference
	{ 	  
	 \{1, \ldots,  n\}\subseteq \contextsTL(\blue{op})\\
	     	} 
	{ \contextsTL \mid \Gamma\typSup, \blue{op} : \text{derived} \typeOf \blue{
           (op [\widetilde{T}] \app V_1\app \cdots \app V_n\app \widetilde{E})\app\stepTL\app e }: \blue{op} : \text{plain}}\quad\textsc{(r-derived)}
\end{gather*}  
}
  \caption{Type system for type soundness: error contexts and reduction rules}
  \label{fig:typeSystemContinued}
\end{figure*}

\section{A Type System for Type Soundness}
\label{sec:typeSystemForTypedLanguages}

In this section, we devise a type system that applies the methodologies 
described in Section \ref{sec:progress:methodology} and \ref{sec:typePreservation}. 
To simplify our presentation we fix, without loss of generality, that the eliminated argument is always the first after the type annotation arguments that the eliminator might have. 
Furthermore, we consider only the case of languages with at most one error. 
Our type system can be generalized easily to the presence of multiple errors. 

The definition of the type system is defined in three figures. 
Figure \ref{fig:typeSystemTL} contains the main type system and the type system for definitions and typing rules. 
Figure \ref{fig:typeSystemContinued} contains the type system for error contexts and for reduction rules. 
Figure \ref{fig:preservation} contains the type system for ensuring the type preservation of reduction rules. 
We begin with the main typing judgment $\typeOf  \Ldl$, for a given typed language $\Ldl$, shown in Figure \ref{fig:typeSystemTL}.

In the first line of premises of \textsc{(ts-main)}, we split the rules of the language into three categories: 
value and error definitions, typing rules and reduction rules. 
Each of these categories is type checked using an appropriate typing judgement. 

The grammar that we employ in our type system is the following. 
Below, $\Gamma$s are type environments as usual, and $B$s stand for bindings. 
\begin{gather*}
\mathcal{X} \in \{\text{d}, \text{t}, \text{r}\}, \noindent c \in \Sigma(\small{\lp{type}}), op \in  \Sigma(\small{\lp{exp}}),
 N \subseteq \N \\
\indent 
\begin{array}{rcl}
\Gamma^{\mathcal{X}} & ::= &\emptyset \mid B^{\mathcal{X}}, \Gamma^{\mathcal{X}}\\ 
B^{\mathcal{X}} & ::= & op : role^{\mathcal{X}}\\
role\defSup & ::= &  \text{value}\app N \mid \text{error}\app N \\
role\typSup & ::= &  \text{value}\app c \app N \mid \text{error}\app N \mid \text{elim}\app c \mid \text{derived}  \mid \text{errorHandler}\\
role\redSup & ::= & \text{plain} \mid \text{eliminates}\app op
\end{array} 
\end{gather*}

The type system $\typeOf\defSub$ type checks value and error definitions and produces \emph{bindings} of type $B\defSup$. 
These bindings simply classify values and errors as such and are collected to form the type environment $\Gamma\defSup$ with $\Gamma\defSup \equalUnique B\defSup_1, \ldots, B\defSup_n$, defined in Figure \ref{fig:typeSystemTL}. $\equalUnique$ collects the bindings and also checks that operators are given a unique role and that there exists only one error operator. 

The type system $\typeOf\typSub$ type checks the typing rules of the language. This type system makes use of $\Gamma\defSup$ and produces bindings of type $B\typSup$. These bindings fully classify all the operators according to the classifications of Section \ref{sec:classification}.  These bindings are collected with $\equalUnique$ in $\Gamma\typSup$. When applied to bindings of type $B\typSup$, $\equalUnique$ makes sure that each operators has only one typing rule. 

The type system $\typeOf\redSub$ type checks the reduction rules of the language and produces bindings of type $B\redSup$. These bindings keep track of the operators that are eliminated by others by means of a reduction rule. These bindings are collected and are checked against the classification in $\Gamma\typSup$ with the operation $\Gamma\typSup \exh (B\redSup_1, \ldots, B\redSup_m)$, defined in Figure \ref{fig:typeSystemTL}. Intuitively, \texttt{exh} stands for \emph{exhaustiveness}. This predicate checks whether each eliminator eliminates \emph{all} the values of the type they eliminate, that the error is eliminated by the error handler, and that the error handler has a reduction rule that fires for values. 
Notice that we do not require the uniqueness conditions of $\equalUnique$ on bindings of type $B\redSup$. Indeed, an eliminator \emph{must} have more bindings for eliminating multiple values when necessary.

In the fifth line of premises of \textsc{(ts-main)} we check the correctness of the evaluation contexts with $\key{well\textendash formed}(\contextsTL)$, which is defined in Figure \ref{fig:typeSystemTL}. This check makes sure that each operator does not have context declarations with circular dependencies, as prescribed by \textbf{D3}. Furthermore, arguments that are tested for valuehood are set as evaluation contexts, as prescribed by \textbf{D1}. 

The fifth line of premises of \textsc{(ts-main)} also handles error contexts with the typing judgement $\contextsTL\mid \Gamma\typSup \typeOf  \errContextsTL$, defined in Figure \ref{fig:typeSystemContinued}. This type system accommodates three cases. When the error is not present at all then the error context must be \text{None}. When the error is present but no error handler is defined then the error contexts must coincide with the evaluation contexts. Ultimately, when the error and the error handler are present we check that the error contexts are the evaluation contexts minus the error handler at the eliminated argument, as prescribed by \textbf{D2}. 

Finally, the last line of premises of \textsc{(ts-main)} makes use of the type system $\typeOf^{\Ldl}\preSub$ for 
checking whether all the reduction rules are type preserving. 

Below, we explain the type systems in detail. In what follows, 
the notation $\widetilde{X}$ is short for $X_1 \cdots X_n$ and denotes a finite number of distinct variables as arguments, e.g., $(\lpsyntax{f}\app \widetilde{X}) \equiv (\lpsyntax{f}\app X_1 \cdots X_n)$. As previously defined, variables $V$ are value variables. To avoid confusion, we use $E$ for expression variables that cannot be value variables and $E\valueSup$ for expression variables that may, or may not, be value variables.

\paragraph{A Type System for Definitions}
The type system for definitions has a judgement of the form $\contextsTL \typeOf\defSub \phi : B\defSup$. 
The context $\contextsTL$ is necessary for checking that progress-dependent arguments of values 
and the error are contextual. 

\textsc{(d-value)} processes a value definition and classifies the operator as value. 
Notice that at this point, we do not know which type the operator builds a value of. 
This information is stored in typing rules and will be added later. 
The type assigned by this meta typing rule keeps the information $N$ of the arguments that need to be values for the definition to apply. 
This information is needed when type checking the reduction rules of eliminators, as explained later. 

\textsc{(d-error)} has the same role as \textsc{(d-value)} but for the error definition. 

\paragraph{A Type System for Typing Rules}
The type system for typing rules has a judgement of the form $D\mid \Gamma\defSup \typeOf\typSub \app \phi : B\typSup $. 
The argument $D$ is the set of reduction rules of the language. 
This argument is needed for distinguishing the role of some operators. 

\textsc{(t-value)} applies to typing rules of operators that $\Gamma\defSup$ classifies as values. 
The shape of the typing rule deserves some attention. This shape imposes that the assigned type have the form $\blue{(c \app \widetilde{T_2})}$, that is, a constructed type. 
Notice that we rely on the first classification with $\typeOf\defSub$ to know whether the operator has been classified as value. In particular, we do not label values only on the ground 
of encountering constructed types as assigned type. 
For example, $\small{\lp{tail}}$ builds an expression of type $\small{\lp{list}}$ 
but it is not a value for lists. 
Similarly, $\small{\lp{isNil}}$ builds boolean expressions but it is not a value of type $\small{\lp{bool}}$. 
Therefore, we look at the classification $\typeOf\defSub$ for help. At this point, \textsc{(t-value)} simply discovers 
the type constructor that the value is associated with and passes this information along. 

Another characteristic to notice on the shape of the typing rule is that it imposes that all the arguments of the operator are the subject of a typing premise, as prescribed by \textbf{P-Typ}. 
Throughout the type system, we fix the convention that $\blue{\lpGamma_1}, \ldots \blue{\lpGamma_n}$ are build with $\blue{\lpGamma}$ and they exclusively can be of the form 
\[
\blue{\lpGamma_i} ::=
\blue{\lpGamma}\mid\blue{\lpGamma}, x\mid\blue{\lpGamma}, x:T \quad (i\in \N)
\] 
This means that \textsc{(t-value)} allows for ordinary typing premises as well as 
generic and hypothetical premises.

\textsc{(t-elim)} classifies eliminators at the encounter of their typing rule. 
The shape of the rule imposes that the type of the eliminated argument has the form $\blue{(c \app \widetilde{T})}$, that is, a constructed type. This is not sufficient for labeling the operator as eliminator. For example, the argument of $\small{\lp{succ}}$ is $\small{\lp{int}}$, constructed type, but $\small{\lp{succ}}$ is a constructor. 
If we, additionally, check the mere presence of reduction rules for the operator at hand, it would not be sufficient either. For example $\small{\lp{fix}}$ has a reduction rule and its only argument is typed at $\small{\lp{arrow}}$, constructed type, but $\small{\lp{fix}}$ is not an eliminator.
Therefore, we check whether a reduction rule for the operator eliminate a value. 
This is done with the auxiliary function $D^{\text{r}}.\field{rules}(\blue{op_1})$ that denotes the set of rules in $D^{\text{r}}$ whose source expression is build with $\blue{op_1}$. We match each reduction rule with the form 
$\small{\blue{(op_1 [\widetilde{T_2}]\app \HI{$(op_2[\widetilde{T_3}]\app \widetilde{E\valueSup_1})$} \app \widetilde{E\valueSup_2}) \app \stepTL\app e }}$. 
Notice that the high-lighted expression is a constructed expression. 
At this point, we check whether $op_2$ has been classified as value. 

\textsc{(t-errHandler)} classifies the error handler in a way that is similar to that of \textsc{(t-elim)}. This time, we check that $op_2$ has been classified as the error. 
%

\textsc{(t-derived)} classifies derived operators. We check that all the reduction rules for the operator are of the form 
$\blue{(op [\widetilde{T_i'}]\app \widetilde{E\valueSup_i}) \app \stepTL\app e_i}$. This means that the arguments of $\blue{op}$ are all variables, whether value or  expression variables. In particular, there is no pattern-matching of constructed expressions. \\ \indent
%
\textsc{(t-error)} handles the typing rule for the error. In this rule, we use $\key{var}(ty)$ to denote the set of variables in $ty$. We enforce that the assigned type is a free variable $T$. 
This makes sure that the error can be typed at any type, as prescribed by \textbf{D-Err}. 

\paragraph{A Type System for Reduction Rules}
The type system for type checking reduction rules has a judgement of the form $\contextsTL \mid \Gamma\typSup \typeOf\redSub \phi:  B\redSup $. The binding produced by this judgement records whether an operator eliminates another one. This happens for reduction rules of eliminators and for the reduction rule that handles the error. In those cases the produced binding has the form $(\blue{op_1} : \text{eliminates } \blue{op_2})$. 
All other reduction rules produce a binding with the label ``plain'', which means that no elimination takes place.
We show this type system in Figure \ref{fig:typeSystemContinued}. 

\textsc{(r-elim)} type checks a reduction rule for an eliminator. The shape of this rule must be of the form $\blue{(op_1 [\widetilde{T_1}] \app
           (op_2 [\widetilde{T_2}] \app \widetilde{V} \app \widetilde{E}) \app \app\widetilde{V'} \app \widetilde{E'})\app\stepTL\app e}$. 
           We check that $\blue{op_1}$ is an eliminator for some type constructor $\blue{c}$ and that $\blue{op_2}$ is a value for that specific type. 
            We impose that the rule fires exactly when $\blue{op_2}$ forms a value. To this aim, $\blue{(op_2 [\widetilde{T_2}] \app \widetilde{V} \app \widetilde{E})}$ must be such that the variables $\widetilde{V}$ are precisely those prescribed by $N$. 
            With the check $\{1, \ldots m\} \subseteq \contextsTL(\blue{op_1})$ we impose that the eliminated argument (index $1$) is contextual, and that also its sibling arguments that are tested for valuehood are. This is prescribed by \textbf{D1}. \\ \indent
\textsc{(r-errHandler)} type checks the reduction rule that handles the error. The way we handle this case is very similar to that of \textsc{(r-elim)}. It differs from \textsc{(r-elim)} in that it makes sure that $\blue{op_1}$ is the error handler and that $\blue{op_2}$ is the error. \\ \indent 
\textsc{(r-errHandler-value)} type checks the reduction rule that defines the step of the error handler for values. 
The form of the rule must be $\blue{
           (op [\widetilde{T}] \app V_1\app \widetilde{V}\app \widetilde{E})\app\stepTL\app e }$. This imposes the eliminated argument to be a value variable. As for \textsc{(r-elim)} and \textsc{(r-errHandler)}, we impose that the eliminated argument (index $1$) and those sibling arguments that are tested for valuehood are contextual. 
           \\ \indent
\textsc{(r-derived)} type checks the reduction rules for derived operators. The shape of these rules imposes that no pattern-matching would take place. As for the previous cases, we then check that evaluation contexts are properly defined. 

\paragraph{A Type System for Type Preservation}
We now explain the type system that ensures that reduction rules are type preserving. 
This type system is presented in Figure \ref{fig:preservation}. 
The judgement for this type system takes the form $D \typeOf^{\Ldl}\preSub \phi$. The argument $D$ is the set of typing rules of the language. Typing rules are necessary because we build symbolic type environments out of them. 
Figure \ref{fig:preservation} shows the type system for $\typeOf^{\Ldl}\preSub$. Given a constructed expression $\blue{e}$, the function $D^{\text{t}}(\blue{e}).\field{premises}$ retrieves the premises of the typing rule of the top level operator of $\blue{e}$ when the rule is instantiated with $\blue{e}$. For example, $D^{\text{t}}(\blue{\small{(\lp{cons} \app V_1\app V_2)}}).\field{premises} = \{\small{\blue{\lpGamma \typeOf \app V_1 : T}}, \small{\blue{\lpGamma \typeOf \app V_2 : \lp{list} \app T}}\}$ (the original rule uses $E_1$ and $E_2$ in lieu of $V_1$ and $V_2$). As we have formed $\Gamma\typSup$ with $\equalUnique$ there is only one typing rule per operator. 
Analogously, we write $D^{\text{t}}(\blue{e}).\field{output}$ for retrieving the assigned type of the typing rule of the top level operator of $\blue{e}$ when the rule is instantiated with $\blue{e}$.
In Figure \ref{fig:preservation}, we also lift the notation $\widetilde{e}$ to expressions with the obvious meaning. 

Rule \textsc{(pre-main)} treats a rule of the form 
$\small{\blue{(op[\widetilde{T}] \app \widetilde{e}\,)\app \stepTL\app e'}}$. 
           Recall that, virtually, we need to establish that $\small{\blue{(op[\widetilde{T}] \app \widetilde{e}\,)}}$ and $\blue{e'}$ have the same type. 
           To do this, we compute the symbolic type environment with the call $D^{\text{t}} \typeOf\chrSub  \small{\blue{(op[\widetilde{T}] \app \widetilde{e}\,)}} : \Gamma^{s}$. 
           The type judgement  $\typeOf\chrSub$ takes an expression and returns a symbolic type environment, 
           that is simply a set of typing formulae. Rule \textsc{(symb-one)} handles the case where $\small{\blue{(op[\widetilde{T}] \app \widetilde{e}\,)= (op [ \widetilde{T}]
	 \app \widetilde{E\valueSup})}}$, that is, all arguments are variables and none are pattern-matched. This happens for reduction rules for derived operators, for example. In this case, we build the symbolic type environment with the premises of the typing rule for $\blue{op}$, suitably instantiated. Reduction rules for eliminators and for handling the error are such that $\small{\blue{(op[\widetilde{T}] \app \widetilde{e}\,) = (op_1 [\widetilde{T}]
          \app  (op_2  [\widetilde{T'}]\app \widetilde{E\valueSup_1})\app \widetilde{E\valueSup_2})}}$, that is, the eliminated argument is built with a top level operator $\blue{op_2}$. This case is handled by \textsc{(symb-two)}, which builds the symbolic type environment with both the typing premises from $\blue{op_1}$ and $\blue{op_2}$, suitably instantiated. 
           Once we have computed the symbolic type environment $\Gamma^{s}$, we check that the source and the target of the reduction rule are typed at the same type when $\Gamma^{s}$ is used. This type is the type assigned by the typing rule of $\blue{op}$ when instantiated to type $\blue{(op[\widetilde{T}] \app \widetilde{e}\,)}$. 
           We check this with $\vdash^{\Ldl}\entSub$, which builds the appropriate query that we check for entailment. 
The function $(\cdot)^\forall$ simply quantifies universally over all the
variables of the query. Notice that the query is checked in the language 
augmented with the axiom for \textsc{(eq-sub)}, which in our setting translates as
\begin{align*}
\textsc{(eq-sub)}^{*} & = \blue{\forall E_1, E_2, T_1, T_2,}\\ 
& \blue{(\forall x.\app \typeofTL \app x : T_1 \Rightarrow \typeofTL \app (E_1\app x) : T_2)
\land \typeofTL \app E_2 : T_1 \Rightarrow \typeofTL \app (E_1\app E_2) : T_2}.
\end{align*}
As $\blue{E_1}$ is an abstraction, $\blue{(E_1\app E_2)}$ encodes the substitution ${E_1[E_2/x]}$ in HOAS.

\begin{figure}
\scriptsize{
\fbox{$ D \typeOf^{\Ldl}\preSub \phi $}
\begin{gather*}
\inference
	{  
              D^{\text{t}} \typeOf\chrSub  \blue{(op[\widetilde{T}] \app \widetilde{e}\,)} : \Gamma^{s}
             	\\
		\blue{ty} = D^{\text{t}}(\blue{(op[\widetilde{T}] \app \widetilde{e}\,)}).\field{output}\\
              \Gamma^{s} \vdash^{\Ldl}\entSub \blue{(op[\widetilde{T}] \app \widetilde{e}\,)}: \blue{ty} &
              \Gamma^{s} \vdash^{\Ldl}\entSub \blue{e'} : \blue{ty}\\
          } 
	{  D^{\text{t}}\typeOf^{\Ldl}\preSub \blue{(op[\widetilde{T}] \app \widetilde{e}\,)\app \stepTL\app e'}}\textsc{(pre-main)}
          \\[1.5ex]
\Gamma^{s}\vdash^{\Ldl}\entSub \blue{e} :
  \blue{ty} \equiv (\Ldl \cup \textsc{(eq-sub)}^{*}) \models \blue{(\Gamma^{s} \Rightarrow \typeofTL
  \app e: ty)}^{\forall}
\end{gather*}  

\fbox{$  D  \typeOf\chrSub \blue{e} : \Gamma^{s} $}
\begin{gather*}
	{  D^{\text{t}}  \typeOf\chrSub \blue{(op [ \widetilde{T}]
	 \app \widetilde{E\valueSup})} :  \bigwedge D^{\text{t}}(\blue{(op [ \widetilde{T}]
	 \app \widetilde{E\valueSup})}).\field{premises}}\quad\textsc{(symb-one)}
\\[1.5ex]
\inference
	{ 
          \Gamma^{s}_1=  \bigwedge D^{\text{t}}(\blue{(op_1 [\widetilde{T}]
          \app  (op_2  [\widetilde{T'}]\app \widetilde{E\valueSup_1})\app \widetilde{E\valueSup_2})}).\field{premises} \\
          \Gamma^{s}_2=  \bigwedge D^{\text{t}}(\blue{(op_2  [\widetilde{T'}]\app \widetilde{E\valueSup_1})}).\field{premises}
        }         
	{  D^{\text{t}}  \typeOf\chrSub \blue{ (op_1 [\widetilde{T}]
          \app  (op_2  [\widetilde{T'}]\app \widetilde{E\valueSup_1})\app \widetilde{E\valueSup_2})}
          : \Gamma^{s}_1 \land \Gamma^{s}_2 }\ \textsc{(symb-two)}
\end{gather*}  
}
 \caption{Type system for ensuring type preservation}
  \label{fig:preservation}
\end{figure}

\section{Well-Typed Languages are Sound}
\label{sec:not-unsound}

We are now ready to establish our main results. 
We rely on the type system of logic programs (in the sense of Church, see \cite{Miller:2012aa}). 
This type system is denoted with $\typeOf_{\text{lp}}$ and rejects ill-typed logic programs with 
mistakes such as $\blue{\typeofTL ~T:T}$ and $\small{\blue{(\lp{app}~\lp{arrow}~\lp{arrow})}}$. 
Thanks to $\typeOf_{\text{lp}}$, our type system $\vdash \Ldl$ does
not check for those errors and could focus on its higher level task.
Below, we use $\typeOf_{\text{lp}}$ lifted to typed languages. 
 \[
 \typeOf\tsSub \Ldl\app \equiv \app\app \typeOf_{\text{lp}}\Ldl 
  \text{  and } \typeOf \Ldl.
 \]
%
%
\begin{theorem}[Well-typed languages afford progress]\label{thm:progress}
For all typed languages $\Ldl$ and for all $\blue{e}$ and $\blue{T}$, if $\typeOf_{\mathrm{ts}}
\Ldl$ and $\Ldl\models \small{\blue{\typeofTL \app e : T}}$ then either
$\Ldl\models \small{\blue{\valueTL \app e}}$, 
$\Ldl\models \small{\blue{\errorTL \app e}}$, or
there exists $\blue{e'}$ such that $\Ldl\models \small{\blue{e\app \stepTL \app e'}}$.
\end{theorem}

\begin{theorem}[Well-typed languages are type preserving]\label{thm:preservation}
For all typed languages $\Ldl$ and for all $\blue{e}$, $\blue{e'}$ and $T$, if $\typeOf_{\mathrm{ts}}
\Ldl$, $\Ldl\models \small{\blue{\typeofTL \app e:
  T}}$ and $\Ldl\models  \small{\blue{e\app \stepTL \app
  e'}}$ then $\Ldl\models \small{\blue{\typeofTL \app e':
  T}}$.
\end{theorem}


Type soundness follows from the progress and
preservation theorems in the usual way. 

\begin{theorem}[Well-typed languages are sound]
For all typed languages $\Ldl$ and for all $\blue{e}$, $\blue{e'}$ and $\blue{T}$, if $\typeOf_{\mathrm{ts}}
\Ldl$, $\Ldl\models \small{\blue{\typeofTL \app e:
  T}}$ and $\Ldl\models \small{\blue{e\app\multistepTL \app
  e'}}$ then either
\begin{itemize}
\item $\Ldl\models \small{\blue{\valueTL \app e'}}$, 
\item $\Ldl\models \small{\blue{\errorTL \app e'}}$, or
\item there exists $\blue{e''}$ such that $\Ldl\models  \small{\blue{e'\app  \stepTL \app e''}}$.
\end{itemize}
\end{theorem}

The proofs of the theorems above can be found in the appendix. 

\section{Implementation: the \emph{\certifier}}
\label{sec:implemenation}
Based on the work of this paper, we have implemented a tool that 
we have called \emph{\certifier}. The tool is written in Ocaml and reads Abella specifications (basically $\lambda$Prolog 
specifications) augmented with special tags for declaratively specifying evaluation contexts.  
The tool implements a type-checker based on the type system of
Sections \ref{sec:typeSystemForTypedLanguages}. 
We have realized the type system for type preservation by automatically generating queries to 
the Abella theorem prover.

We have applied
our tool to several variants of the simply typed lambda calculus with various subsets 
of the following features: pairs,
\key{if\textendash then\textendash else}, lists, sums, unit, tuples,
\key{fix}, \key{let}, \key{letrec}, universal types, recursive types
and exceptions. We have also considered different strategies such as call-by-value, call-by-name
and a parallel reduction strategy, as well as lazy pairs, lazy lists
and lazy tuples.
We have type checked a total of $103$ type sound languages, including a rich language such as $\lambdafull$. 

Remarkably, \emph{\certifier} spots design mistakes that hinder type soundness. Among other kinds of errors, the tool pinpoints the cases when 
\begin{itemize}
\item Some eliminator does not eliminate all the values it is supposed to eliminate. 
\item Some relevant evaluation context is not declared.
\item Context declarations have circular dependencies such as $\key{cons}\app E\app v\mid\key{cons}\app v\app E$, mentioned in Section \ref{circular}. 
\item (\redd{\#}) Some reduction rules are not type preserving. For example, 
if we mistake the operational semantics of \key{head} and define it to
  return the second component, that is, the rest of the list, the type-checker 
  points out the bad rule. We will refer to this item as (\redd{\#}) when speaking of related work. 
\end{itemize}
In general, thanks to our type system setting the tool can algorithmically detect departures from the 
methodology of Section \ref{sec:progress:methodology} and report them to the user. 

\textbf{Certified languages:} For those language specifications that successfully pass our
type checker, \emph{\certifier} automatically produces a
formal proof of type soundness and related theorems. 
These proofs are independently machine-checked by the 
Abella theorem prover \cite{baelde14jfr}. 

A serious investigation on our automatic certification algorithms 
is part of our future work. 

The \emph{\certifier} tool can be found at the following repository:

\centerline{{\texttt{https://github.com/mcimini/TypeSoundnessCertifier}}}


\section{Related Work}
\label{sec:related}


The meta-theory set forth in this paper is inspired by a line of
research on the meta-theory of operational semantics, and especially
on results on \emph{rule formats} \cite{Mousavi:2007}. These results
 offer templates and restrictions to operational
semantics specifications that can guarantee that some property
holds. Typical work from this line of research have
been used for establishing various results for process
algebras and mostly in the context of equations modulo
bisimilarity and congruence \cite{Bloom:1995,CranenMR08,coomm,Aceto:2009}. 
This paper shares the same
strive to ensure properties \emph{by design} for languages given as input. 
However, our results target 
programming languages with types, ensure type soundness, 
and aims at offering a typing discipline rather than syntactic restrictions. 


The specific use of logic programs for
encoding operational semantics and typing rules dates back to
Kahn's \emph{natural semantics} \cite{kahn87stacs} and its machine
implementation \cite{borras88}.  The use of higher-order logic programming
as a specification language dates back to 
Burstall \& Honsell \cite{burstall88} and
Hannan \& Miller \cite{hannan90lfp}.

Automated proving has been explored in the context of type soundness. 
The seminal work of Sch\"urmann and Pfenning shows that important aspects of the meta-theory of programming languages are in the reach of automatic theorem proving in the context of the logic programming based theorem prover Twelf \cite{Pfenning1999,twelf}. 
Their system can establish the type soundness for non-trivial functional languages in a completely automatic fashion, and can do so at the level of machine-checked proofs. 
Similarly, proof assistants such as Coq allow for tactic languages that can automate sophisticated proof patterns. Some well-thought out proof scripts are capable of automating proofs of the progress theorem for some basic languages\footnote{Perhaps, a good example of this is shown in Adam Chlipala's 4-th lecture at the Oregon Programming Languages Summer School 2015 \cite{videoChlipala}.}. 

In this respect, our results offer a type system based companion analysis technique for the type soundness of languages. 
The analogy with programs is immediate, where automated proving is certainly not the only kind of analysis technique available. Type systems are another important one. To make an example, the existence of automated tools for, say, showing data race freedom (DRF) of programs, do not invalidate the benefits of type systems for DRF. The two analysis techniques simply accompany each together.

The way our tool checks for type preservation is essentially similar to the much earlier work in Twelf. In this regard, the ability of spotting errors of the kind (\redd{\#}) (previous section) is not a novelty for the class of languages we capture. 
On the other hand, the way we type check for the progress theorem and report language-specific errors seems to be a novelty in this area. 

There are several tools that support the
specification of languages, such as Ott \cite{Sewell:2007}, Lem \cite{Mulligan:2014}, the K
framework \cite{rosu-serbanuta-2010-jlap}, and PLT
Redex \cite{redex}, among others. In many ways, \emph{\certifier} shares with them the same
spirit in assisting language designers with their designs. 
To our knowledge, the use of a type checker over language definitions and the way the tool informs language 
designers of design mistakes w.r.t. soundness are novelties in tools for language design. In this respect, 
\emph{\certifier} presents features that are orthogonal to those of the mentioned tools. 
Of course, these other mature tools offer remarkable help to language designers
in multiple aspects, including features for executing, evaluating, testing and exporting
language specifications.

\section{Conclusions and Future Work}
\label{sec:conclusion}

In this paper, we somehow treated language specifications
as expressions and we have demonstrated that the appropriate typing discipline over 
these specifications 
guarantees that the language is type sound: that is, \emph{well-typed
  languages are sound}. 

We have demonstrated this idea with a class of languages based on constructors/eliminators and 
errors/error handlers: features that are common in programming language design. 
This class is fairly expressive and comprises languages with modern features such as recursive types, polymorphism and exceptions. 

Are there programming languages that are out of the reach of our results?
Yes, definitely many.  This is our first paper on the topic and we have only scratched 
the surface of this research area. 
Perhaps, the two most natural extensions to the present work are to languages with stores/references 
and languages with subtyping. 
These extensions are not as trivial as they might seem. 
For example, languages with stores/reference carry a heap, and a notion of safety must be systematically derived for the heap. These languages also impose adjustments to the preservation theorem statement for accommodating a location environment that might grow over time.

Similarly, languages with subtyping bring their own difficulties. For example, as both the language and the subtyping relation are provided by the language designer, we would need principled ways to enforce that object subtyping is rejected when it is covariant in calculi with updates (unsound), as well as when references are covariant (unsound), and all similar scenarios. We leave an investigation of these classes of languages as future work. 

Other classes of languages, such as linear types, dependent types, type-effect systems and typestate, to name a few, are  out of the scope of our type system and they seem to come with their own domain-specific research challenges. We leave these extensions to future work. Similarly, we plan to investigate whether we can translate our results to the style of big step operational semantics.

In this paper, we conjecture that ``\emph{well-typed languages are sound}'' is a perspective that, 
just like ``\emph{well-typed programs cannot go wrong}'', applies across several classes of languages. 
We will be eager to work with the community to explore this research area further.  
%
%


%

\bibliographystyle{plain}
\bibliography{all,local}

\appendix

\section{Progress Theorem}

\textbf{Remark: } 
The proofs in Appendix A, B and C suggest an algorithm for producing the theorems (which are language-dependent) and the proofs that are related to the type soundness proof. These algorithms have been implemented in the \emph{TypeSoundnessCertifier}. Spelling out the algorithms and embarking on a serious account of them will be part of a subsequent paper.

\paragraph{The Main Progress Theorem}

Assume $\typeOf\tsSub \Ldl$ and $\Ldl\models \typeofTL \app e : T$. 
The proof is by induction on $\Ldl\models \typeofTL \app e : T$. 
Being $\Ldl\models \typeofTL \app e : T$ provable, it means that there exist a typing rule $\phi$ of the form 
$\inference{p_1, \ldots p_n}{\typeofTL\app (op[\widetilde{T}]\app \widetilde{E}):t}$ that is 'is satisfied'. 
The rule 'is satisfied' in the sense that there exists a substitution $\gamma$ from logical variables (of the rule) to logical terms such that $\Ldl\models p_i\gamma$ for $\{1, \dots, n\} $ and $\typeofTL\app (op[\widetilde{T}]\app \widetilde{E})\gamma:t\gamma =  \typeofTL \app e : T$.

Since $\phi\in \Ldl$ and $\phi$ is a typing rule, then it has been type checked with $\typeOf\typSub$. 
This means that \emph{all} variables in $\widetilde{E}$ are the subject of a typing premise (\textbf{P-Typ} common pattern), 
i.e. $\Ldl\models \typeofTL \app E_i\gamma \app T\gamma$ for $E_i \in \widetilde{E}$. This means that we can apply the inductive hypothesis to each $E_i$ if we wish. Of course, it matters to apply the inductive hypothesis to progress-dependent arguments only, if we were to be optimal. The paper does not set a notation for extracting progress-depending arguments, we simply apply the inductive hypothesis to the contextual arguments of $op$. This is suboptimal (only slightly) but correct. 

Notice also that in HOAS some variables might be abstractions and might be subject to typing premises $p_i$ that might be hypothetical or generic. However, the shape of the value premises, value definitions and error definitions is of the simple form $\valueTL\app V$: this implicitly forbids  evaluation under a binder because to define that we need a generic premise that wraps a value premise (evaluation under binders is not common in programming languages). In short, the contextual variables are of simple expression variables and we can apply IH as usual. We retrieve the contextual arguments and proceed in the following way. 

By definition of $\Ldl$, $\Ldl$ has the function $\contextsTL$ for contexts. 
Given the operator $\op$ above, and given $\{i_1, \dots, i_n\} \in \key{fst}( \contextsTL({op}))$, we apply the inductive hypothesis to $\Ldl\models \typeofTL \app E_{i_j}\gamma \app T\gamma$, for all $1 \leq j \leq n$. 
Now we have that those $E_{i_j}\gamma$ progress. We call the Progress Lemma for $op$ (defined below) passing the assumptions that $E_{i_j}\gamma$ progress.
Notice that such lemma expects exactly those progress assumptions and in that number (the number of contextual arguments), 
as explained below. 

\paragraph{Progress Lemma for all $\op$}


Given an operator $op$ of kind $(\ldots \rightarrow \lp{term})$ in $\Ldl$, we prove the following theorem.
\begin{theorem}
if $\typeOf\tsSub \Ldl$, with $\Sigma$ being the signature of $\Ldl$,  it holds that 
for all $\op \in \Sigma({exps})$, for all $\{i_1, \dots, i_n\} \in \key{fst}( \contextsTL(op))$, if 
progress $e_1$, $\ldots$
progress $e_n$, then progress $(\op\app e_1\ldots \app e_n\app \widetilde{e})$ for all $e_1$,  $\ldots$, $e_n$ and $\widetilde{e}$ (here $\widetilde{e}$ are the rest of the arguments, respecting the arity of {op}, that are not contextual). 
\end{theorem}

The proof is by case analysis on all progress $e_1$, $\ldots$ progress $e_n$, but in a suitable order.
Since $\typeOf\tsSub \Ldl$ we have that $\contextsTL(op)$ does not declare acyclic dependencies, therefore we can choose an order such that (\textbf{invariant:}) we do case analysis on progress $e_i$ before the case analysis on progress $e_j$ if the context for $i$-th argument of $op$ does not depend on the valuehood of the argument $j$ of $op$. 

After the series of cases analysis on progress $e_i$, we are at the leftmost child of the leftmost tree of the cases. 

\emph{Before continuing: an example.} If we have two arguments, after the first case analysis on $progress\app e_1$ we open three cases: 
1) $value \app e_1$ and $progress \app e_2$, 2) $step \app e_1$ and $progress \app e_2$, 3) $error \app e_1$ and $progress\app e_2$. We then are at the left child. 
We now do case analysis on  $progress \app e_2$ and we open other three cases only on the left child: the leftmost subtree is
1) $value\app e_1$ and $value \app e_2$, 2) $value\app e_1$ and $step \app e_2$, 3) $value \app e_1$ and $error \app e_2$. And we are at the leftmost child: $value \app e_1$ and $value \app e_2$. 

Now we continue the proof. After the series of cases analysis on progress $e_i$, we are at the leftmost child of the leftmost tree of the cases. In this case, \emph{all arguments are values}. 
 
The proof is by case analysis on how $\op$ has been classified with $\typeOf\typSub$.

\begin{itemize}
\item $op : \text{value}\app c \app N$: We dismiss the leftmost child in the following way: Since the typing rule $\phi$ has been typed, then $\Gamma\defSub$ contains $op : \text{value} \app N$. Which means that there exists a value definition $\phi^{d}$ for $op$. Since $\phi^{d}$ has been typed with $\Gamma\defSub$, it means that the shape of the value deifinition is such that it is restricted only by value premises, i.e. by the valuehood of its arguments (this realizes the common pattern \textbf{P-Val}). As we are in the case where all the arguments are values, the definition applies and this case progresses. 

Now, we are left with two cases: 1) all arguments are values but the last one which is $step \app e_n$ and 2) all arguments are values but the last one which is $error \app e_n$. We can treat these two cases uniformly for all the tree that the case analysis generated. Indeed, notice that once we dismiss these two cases we have dismissed the whole case $value$ of the subtree immediately above. Therefore, we go straight to prove the cases $step$ and $error$ of the subtree immediately above. 
We can use the invariants on the dependency on the valuehood for proving all those cases in the same way at any level of the tree. We have
\begin{itemize} 
\item \textbf{STEP:} $step \app e_j$ i.e. for some $j$. As $j \in  \contextsTL({op})$ by the semantics of $\Ldl$ (translation to logic programs), this means that there exists a rule \\
 $\inference
{\widetilde{\valueTL\app E} &  \app E_j\app \stepTL \app E_j'}
{ ({op}[\widetilde{T}]\app  \cdots E_j\cdots )\app\stepTL \app ({op}[\widetilde{T}]\app  \cdots E_j'\cdots )}$. Notice that we have ordered the arguments by dependency on valuehood, therefore value premises can be applied, if any, only to $E_1 \ldots$ previous to $E_j$. 
However, by the invariant that we get from acyclic contexts, we could chose an order that deals with the case $step \app e_j$ only when the previous arguments are values. So we can instantiate and prove a step  $$\Ldl\models { ({op}[\widetilde{T}]\app  \cdots e_j\cdots )\app\stepTL \app ({op}[\widetilde{T}]\app  \cdots e_j'\cdots )}$$ 
So $(op[\widetilde{T}]\app \widetilde{E})$ progresses.

\item \textbf{ERR:} $error \app e_j$ for some $j$, i.e. $e_j$ is an error. Since $j \in  \contextsTL({op})$ and since $\op$ is not an error-handler then $j \in  {err\textendash ctx}({op})$. By by the semantics of $\Ldl$ (translation to logic programs) this means that there exists a rule  $\inference
{\widetilde{\valueTL\app E} & \errorTL \app E_j}
{ ({op}[\widetilde{T}]\app  \cdots E_j\cdots )\app\stepTL \app E_i}$. Again, as we have ordered the argument by dependency on valuehood, the arguments $e_1 \ldots$ are values and the rule can be applied to prove the step $\Ldl\models  ({op}[\widetilde{T}] \cdots \app e_j\app\cdots) \app\stepTL \app e_j$. So $({op}[\widetilde{T}]\app \widetilde{E})$ progresses.
\end{itemize}

\item $op : \text{elim}\app c$: As $\phi$ is a typing rule of $\Ldl$ and $\vdash  \Ldl$, we have that $\phi$ has been type checked by $\vdash\typSub$. This means that $\phi$ is of the following shape. 
\[r= \inference
	{\typeofTL \app E_1:  (c\app \widetilde{T})} 
	{\typeofTL\app  ({op}[\widetilde{T}]\app  \widetilde{E}):ty}\]
	
	Since rule $r$ has been satisfied, so are its premises. Then, we have $\Ldl\models \typeofTL \app e_1:  (c\app \widetilde{T})$. Since we are in the leftmost case, where all arguments are values, we have $\Ldl\models \valueTL \app e_1$. Therefore, we apply the Canonical Forms Lemma for $c$ (described in the following paragraph). This means that $e_1 = \{(t_1, \valuesTL_1) \lor \ldots \lor (t_m, \valuesTL_m)\}$ (this is notation from the next paragraph). Let us fix one such pair $(t_k, \valuesTL_k)$. By $\typeOf \Ldl$, we have $t_k = ({op_2}[\widetilde{T}]\app \widetilde{E'})$. This means that $\vdash\typSub$ has type checked $op_2$ as $op_2 : \text{value} \app c\app \valuesTL_k$. Since $\vdash\Ldl$ succeded also $\vdash\redSub$ succeded, which means that  the exhaustiveness check $\Gamma\typSup - (B\redSup_1, \ldots, B\redSup_m) = \emptyset$ succeded, and means that since $op_2 : \text{value} \app c\app \valuesTL_k \in\Gamma\typSub$ then we had a reduction rule $r_{\text{step}}$ such that has been type checked by $\vdash\redSub$ as $op :\app\text{eliminates}\app op_2$, because $op : \text{elim}\app c$. 
	 
	Since $r_{\text{step}}$ has been type checked by $\vdash\redSub$ with \textsc{(r-elim)} it is of them form: 
	$r_{\text{step}} = \inference{ps}{ (op[\widetilde{T}] \app (op_2[\widetilde{T}]\app \widetilde{E\valueSup_1}) \app \widetilde{E\valueSup_2})\app\stepTL\app exp}$. Therefore we could apply this rule, only provided that premises $ps$ are satisfied. The shape of the rule also imposes that $ps$ are only value premises. 
	These premises are of two kind: 
\begin{itemize}
	\item $\valueTL \app E_u$ where $E_u \in \widetilde{E\valueSup_2}$. Then we are in the case where all of those $E$s are values. Indeed, those arguments are progress-depending arguments for needing valuehood. Also, we are in the leftmost case of the case analysis on all progresses $e_j$ on progress-depending arguments. Thus, we have $\Ldl\models \valueTL\app  e_u$ for each of them, which satisfies the premise. 
	\item $\valueTL\app  E_u$ where $E_u \in \widetilde{E\valueSup_1}$. Then $\typeOf\prgSub \Ldl$ imposes that the index $u\in \valuesTL_k$, which means $\Ldl\models \valueTL\app  e_u$ (as defined in the next paragraph), therefore also this premise is satisfied. 
		\end{itemize}
		We can therefore apply the rule $r_{\text{step}}$ above and prove $\Ldl\models  (\op \ldots \app e_j\app\ldots) \app\stepTL\app e'$ for some $e'$. So this cases progresses.

		Cases \textbf{STEP} and \textbf{ERR} are proved as in the previous case for constructors. \\

The other operators are easier to handle.

\item $op: \text{error}\app N$: The leftmost leaf of errors is handled similarly as to values and so are \textbf{STEP} and \textbf{ERR}. 
\item $op : \text{derived}$:  Then $\phi$ is typed by $\vdash\typSub$ by \textsc{(t-derived)}. Therefore, it exists a rule $\blue{(op \app \widetilde{V_i} \app \widetilde{E_i'}) \app \stepTL\app e_i}$. As those $V_i$ are tested for valuehood they are progress-dependent arguments, so we are in the case analysis of their progress and in particular, they are all values because we are in the leftmost case. Therefore that reduction rule applies and this case progresses. 
For derived operators, \textbf{STEP} and \textbf{ERR} also follow the same line as in the previous cases. 
\item $op_1 :  \text{errHandler}$: Then $\phi$ is typed by $\vdash\typSub$ by \textsc{(t-errHandler)}. Therefore, it exists a rule 
$\phi_2 = \blue{(op_1 \app V \app \widetilde{E_4}) \app \stepTL\app e_2}$. As we are in the leftmost case, the eliminating argument, i.e. the first argument, is a value and thus we can apply the reduction rule. So this case progresses. 
For error handlers, \textbf{STEP} follows the line as in the other case, while \textbf{ERR} is different: since $\phi$ is typed by $\vdash\typSub$ by \textsc{(t-errHandler)}, it means it exists a rule $\phi_1 = \blue{(op_1 \app
           (op_2\app \widetilde{E_2}) \app \widetilde{E_3}) \app \stepTL\app e_1}$ and 
           $\Gamma\defSup(op_2) = \text{error}\app N$ .Also, $\phi_1$ has been typechecked by $\vdash\redSub$ which ensures that it fires when the first argument is an error. Therefore we can apply this rule. So this case progresses. 
\end{itemize}

\paragraph{Canonical Forms Lemma for $c$}

\begin{theorem}
For all $e$, $c$, if $\typeOf \Ldl$ and $\Ldl\models \typeofTL \app e : (c\app \widetilde{T})$ and $\Ldl\models \valueTL \app e$ then $e = \{(t_1, \valuesTL_1) \lor \ldots \lor (t_m, \valuesTL_m)\}$ where 
 for all $1 \leq j \leq m$, $t_j = \op \app \widetilde{e}$ and 
\begin{itemize}
\item (Part 1) $op : \text{value } c\app V_j\in\Gamma\typSub$.
\item (Part 2) $\Ldl\models \valueTL \app e_i$ when $i\in \valuesTL_j$.
\end{itemize}
\end{theorem}

Proof. Assume the hypothesis. As $\Ldl\models \typeofTL \app e \app (c\app \widetilde{T})$ and $\typeOf
 \Ldl$, then it means that $e$ is typed with a typing rule $r$ whose input is $({op}[\widetilde{T'}]\app \widetilde{E})$, that is, $e =  ({op}[\widetilde{T'}]\app  \widetilde{e})$. 
 
 Part 1: Since $\Ldl\models \valueTL \app e$,  $op:\text{value } V_j\in \Gamma\defSub$, therefore \textsc{t-elim} finds $op:\text{value } V_j\in \Gamma\defSub$ and the typing rule $r$ and classifies  $op:\text{value }\app c\app  $, which means  $op:\text{value } V_j\in \Gamma\typSub$.
 
Part 2: Since we have $\Ldl\models \valueTL \app e$ (recall $e=t_j = \op \app \widetilde{e}$), we have a rule of form 
$\inference{\valueTL\app E_1 \ldots \valueTL\app E_n}
{\valueTL \app (\op[\widetilde{T}] \app E_1\app\cdots \app E_n \app\cdots)}$. Therefore, all $e_i$, ..., $e_n \in \widetilde{e}$ are such that 
 $\Ldl\models \valueTL \app e_i$ for $1 \leq j \leq n$. Now, by \textsc{(value)}, all indexes in $V_j$ are exactly those indexes of the arguments tested for valuehood in that rule. 
  
\section{Type Preservation}

The proof is by induction on $\Ldl\models  e\app\stepTL \app
  e'$.  


As the formula $\Ldl\models  e\app\stepTL \app
  e'$ is provable, it means that there exists a rule of $\Ldl$ that is satisfied and proves the conclusion 
  $e\app\stepTL \app e'$. This rule can have three different shapes: 
    
  \begin{itemize}
  \item \textbf{contextual rule:} $e = (op[\widetilde{T}]\app \widetilde{e})$ and the rule is of the form $$\inference
{\ E_i\app \stepTL \app E_i'}
{ ({op}[\widetilde{T}]\app  \ldots E_i\ldots )\app\stepTL \app ({op}[\widetilde{T}]\app\ldots E_i'\ldots )}$$ 

By the assumptions of the preservation theorem, we have $\Ldl\models \typeofTL \app (op[\widetilde{T}]\app \widetilde{e}):
  T$, and so we have typing rule $\phi$ in $\typedLanguage$ that proves this typeability fact. Also, since $\vdash \Ldl$, we have that $\phi$ is typed by $\typeOf\typSub$. This means that the shape of the rule is such that all arguments $\widetilde{e}$ are typed, including $e_i$, that is 
$\Ldl\models \typeofTL \app e_i\app  T'$. 
As the reduction rule has been satisfied we also have $\Ldl\models  e_i\app\stepTL \app  e_i'$. 
As in the proof for progress, since $E_i$ is a contextual argument it cannot be an abstraction but is a simple expression variable. Then we can apply the inductive hypothesis on it and obtain that $\Ldl\models \typeofTL \app e_i' : T'$.
  It is easy to see that if $\Ldl\models \typeofTL \app (op[\widetilde{T}]\app \widetilde{e}):
  T$ then $\Ldl\models \typeofTL \app (op[\widetilde{T}]\app \widetilde{e}[e_i'/e_i]):
  T$. That is, the reduction is type preserving.  

  \item \textbf{error steps:} $e = (op[\widetilde{T}]\app \widetilde{e})$, $i \in  {err\textendash ctx}({op})$ and the step has been proved with a rule of the form
  $\inference
{\errorTL \app E_i}
{({op}[\widetilde{T}]\app  \widetilde{E})\app\stepTL \app  E_i}
$
The fact that $\typeOf \Ldl$ imposes that there exists a typing rule that types the error. This is because $op : \text{errHandler} \in \Gamma\typSup$ and $\contextsTL = \errContextsTL \,\cup \,\{op \mapsto (1, \emptyset)\}$. And successively, we have that $\typeOf\redSub$ imposes that the a reduction rule for $op : \text{errHandler}$ is well-typed and that consumes the error in $\Gamma\typSub$, which exists only when a typing rule for the error has been type checked. Now, as this rule has been type checked by $\Gamma\typSub$ and by \textsc{(t-error)}, we have that the shape of the rule is such that the assigned type is a fresh new variable. So we can prove $\Ldl\models \typeofTL \app e_i : T$. That is, the reduction is type preserving.
 
  \item \textbf{by reduction rules:}
  We see solely the case for a step of an eliminator. This proof case subsumes that of other reducers (derived operators and error handlers). Assume $e = (op[\widetilde{T}] \app (op_2[\widetilde{T'}]\app \widetilde{e'})\app \widetilde{e''})$ of type $T$ and the following reduction rule by which the step has been proved. 
 $$\inference{ps}{ (op[\widetilde{T}] \app (op_2[\widetilde{T'}]\app \widetilde{E\valueSup_1}) \app\stepTL \app\widetilde{E\valueSup_2})\app\app exp)}$$
 
(As the rule above has been type cheked by $\typeOf\redSub$, $\mathit{ps}$ contains only value premises.)
By the assumptions of the preservation we have that \\
$\Ldl\models \typeofTL \app (op[\widetilde{T}] \app (op_2[\widetilde{T'}]\app \widetilde{e'})\app \widetilde{e''})\app
  T$, then this latter fact is proved with a rule for which there exists a substitution $ \gamma $ that satisfies the rule and such that $(op[\widetilde{T}] \app (op_2[\widetilde{T'}]\app \widetilde{E'}) \app \widetilde{E''}) \gamma = e$. We have to prove that $(exp)\gamma$ is of type $T$.  
Since this is a typing rule of $\Ldl$, we have that it has been typechecked with $\typeOf\typSub$, which means all arguments $\widetilde{e''}$ are well-typed, and also $(op_2[\widetilde{T'}]\app \widetilde{e'})$ is well-typed. 
Now, these expressions are well-typed with a corresponding typing formula $\Ldl\models \typeofTL \app E_i'$ for $E_i'\in \widetilde{E'}$ or 
  $\Ldl\models \typeofTL \app E_i''$ for $E_i''\in \widetilde{E''}$ (we will consider abstractions later). 
  Since we have that the rule has been typechecked with $\typeOf^{\Ldl}\preSub$, we have that those facts have been put in a conjunction $\Gamma^{s}$ and succeed to prove a query that $\Gamma^{s} \vdash (op[\widetilde{T}] \app (op_2[\widetilde{T'}]\app \widetilde{E'}) \app \widetilde{E''}) : T^s$ and also $\Gamma^{s} \vdash  exp : T^s$, where $T^s$ is the type assigned to the entire expression by the typing rule of $op$, and uses variables, hence the $s$ superscript to remind that it is symbolic. 
  Since this query has been checked with universal quantifications over the variables of the query, any instantiations can be concluded. Therefore, we sure had $\Ldl\models \typeofTL \app (op[\widetilde{T}] \app (op_2[\widetilde{T'}]\app \widetilde{E'}) \app \widetilde{E''})\gamma : T^s\gamma$, that is $\Ldl\models \typeofTL \app (op[\widetilde{T}] \app (op_2[\widetilde{T'}]\app \widetilde{e'}) \app \widetilde{e''}) : T$, and we can also conclude $\Ldl\models \typeofTL \app (exp)\gamma : T$. That is, the reduction is type preserving. 
  In our setting of logic programs, some arguments of $\op$ might be abstraction. In that case the query is hypothetical of the form $ \typeofTL \app x : T_1 \Rightarrow \typeofTL \app R : T_2 $, for some $R$ argument of $\op$. Now, there are two cases: either 1) $t$ contains $R$ simply as a variable, i.e. the step simply inherits $R$ as it is, or 2) $t$ contains $(R\app t')$ for some term $t'$, i.e. the step applies a substitution. In case 1) we have that $R$ will have the same type as in $(op[\widetilde{T}] \app (op_2[\widetilde{T'}]\app \widetilde{E'}) \app \widetilde{E''})$ and the query has checked that the whole resulting term turns out to be type preserving. In the second case, the fact that the query has been checked with the axiom \textsc{(eq-sub)}, guarantees us that $(R\app t')$ matches the expected type as well, and, again, any instantiations of $R$ and $E$ will do as well.  

\end{itemize}

\section{Type soundness}

Type soundness of well-typed languages follows from progress and
preservation in the usual way. 


\end{document}